\documentclass[11pt]{article}
\pdfoutput=1
\usepackage{fullpage}

\usepackage{url}
\usepackage{algorithm2e}
\usepackage{enumerate,color,graphicx}
\usepackage{array}
\usepackage[fleqn]{amsmath}

\def\math#1{$#1$}

\def\mand#1{$$#1$$}

\def\frac#1#2{{#1\over #2}}

\def\eqan#1{\begin{eqnarray*}
#1
\end{eqnarray*}}

\DeclareSymbolFont{AMSb}{U}{msb}{m}{n}
\DeclareMathSymbol{\N}{\mathbin}{AMSb}{"4E}
\DeclareMathSymbol{\Z}{\mathbin}{AMSb}{"5A}
\DeclareMathSymbol{\R}{\mathbin}{AMSb}{"52}
\DeclareMathSymbol{\Q}{\mathbin}{AMSb}{"51}
\DeclareMathSymbol{\I}{\mathbin}{AMSb}{"49}
\DeclareMathSymbol{\C}{\mathbin}{AMSb}{"43}

\def\ceil#1{{\left\lceil\,#1\,\right\rceil}}

\def\dotfil{\leaders\hbox to 1.5mm{.}\hfill}
\newcounter{rmnum}
\def\RN#1{\setcounter{rmnum}{#1}\uppercase\expandafter{\romannumeral\value{rmnum}}}
\def\rn#1{\setcounter{rmnum}{#1}\expandafter{\romannumeral\value{rmnum}}}

\begin{document}

\title{Comparing Prediction Market Structures, With an Application to Market Making}
\author{Aseem Brahma, Sanmay Das, and Malik Magdon-Ismail\\
  Dept. of Computer Science\\
  Rensselaer Polytechnic Institute\\
  Troy, NY 12180, USA}
\date{}

\maketitle
\begin{abstract}
  Ensuring sufficient liquidity is one of the key challenges for
  designers of prediction markets. Various market making algorithms
  have been proposed in the literature and deployed in practice, but
  there has been little effort to evaluate their benefits and
  disadvantages in a systematic manner. We introduce a novel
  experimental design for comparing market structures in live trading
  that ensures fair comparison between two different microstructures
  with the \emph{same} trading population. Participants trade on
  outcomes related to a two-dimensional random walk that they observe
  on their computer screens. They can simultaneously trade in two
  markets, corresponding to the independent horizontal and vertical
  random walks. We use this experimental design to compare the popular
  inventory-based logarithmic market scoring rule (LMSR) market maker
  and a new information based Bayesian market maker (BMM).  Our
  experiments reveal that BMM can offer significant benefits in terms
  of price stability and expected loss when controlling for liquidity;
  the caveat is that, unlike LMSR, BMM does not guarantee bounded
  loss. Our investigation also elucidates some general properties of
  market makers in prediction markets. In particular, there is an
  inherent tradeoff between adaptability to market shocks and
  convergence during market equilibrium.

\end{abstract}

\section{Introduction}
\label{section:intro}
Interest in prediction markets has increased significantly in recent
years across academia, policy makers, and the private sector
\cite{WolZit04,berg2003prediction,servan2004prediction,arrow-statement,chen2007utility}.
Wolfers and Zitzewitz \cite{wolfers-five} discuss how prediction
markets have gone from minor novelties to serious platforms that can
have substantial impact on policy and decision-making; prediction
markets are now accepted as information aggregators that produce
quantitative forecasts. Companies like Google,
Microsoft, and HP have deployed prediction markets internally for
forecasting product launch dates and gross sales. Prediction markets
have often outperformed opinion polling: for example, the Iowa
Electronic Markets have usually outperformed opinion polling in
predicting US political races \cite{berg2001results}. There is little
doubt that prediction markets are valuable for information
aggregation for two reasons: (1) They produce meaningful
quantitative forecasts; (2) Those who possess information are better
incentivized and held accountable than they are in alternative
information-gathering methods like surveys or polls. 

Wolfers and Zitzewitz identify five key challenges to the success of
prediction markets \cite{wolfers-five}. First among these is liquidity
provision --
can prediction markets attract sufficient uninformed trading
 to be liquid and attractive to those with information?
Liquidity is the classic chicken-and-egg problem, 
in which some liquidity begets more liquidity. 
Historically, financial markets have often used 
market makers to provide initial liquidity 
to get the ball rolling; financial 
exchanges often provide specific incentives for firms to become 
market makers. Prediction markets have adopted the same idea.
Typically, in prediction markets, the 
market maker is allowed to take on a loss, subsidizing the market, 
to facilitate more liquidity and faster price discovery; this loss 
is taken as a cost of operation. Robin Hanson suggested a family of 
inventory based market makers based on market scoring rules. 
Of these, the one based on the logarithmic market scoring rule 
(the LMSR market maker) is now the \emph{de facto} standard for 
subsidized prediction markets \cite{Han07}. If the main purpose 
of the prediction market is not commercial, loss-making 
market makers can make a lot of sense, but as prediction markets 
become less experimental, the subsidies become a real loss which
must be  minimized.

The LMSR market maker is appealing on several levels: (1) it has a
strong guarantee on the amount of loss it can suffer; (2) since it is
purely inventory based, it is difficult to manipulate in some settings
\cite{chen2009gaming}; (3) it can be shown that, under certain
conditions, particularly that 
participants are rational and learn from prices, a market mediated by
an LMSR market maker will converge to the rational expectations
equilibrium price \cite{PenSam07}.  However, the LMSR also suffers from
several drawbacks, some serious: (1) Although it is bounded, the
market maker does typically run at a loss, which can be large, (2) A
single parameter, $b$, controls many different aspects of the market
maker's behavior, including the loss bound, the level of liquidity in
the market, and the rate of adaptivity to market shocks; setting $b$
to optimally manage these tradeoffs is considered something of a ``black art''
\cite{DBLP:conf/sigecom/OthmanSPR10};  (3) when the posterior belief of the trading population
does not converge (which is likely when people have independent
information and valuations), the price does
not converge to a well defined probability estimate, instead
fluctuating about the equilibrium price; the fluctuations are
asymmetric and again sensitive to the choice of $b$, making it difficult to
extract a quantitative probability estimate; (4) The market maker
provides only point probabilities over outcomes and cannot be easily
coupled with a measure of uncertainty; (5) the market maker
cannot easily be applied to unbounded markets.

An alternative to inventory based market makers is an 
information based market maker. 
The seminal paper of Glosten and
Milgrom \cite{GloMil85} introduced a model of market making
under asymmetric information. Building upon this model, Das \cite{Das05a,Das08} 
and Das
and Magdon-Ismail \cite{DasMag08}
have described efficient market making algorithms for 
zero-profit (competitive) and profit maximizing (monopolist) market makers.
These market makers address some of the drawbacks of the 
LMSR market maker. Specifically, 
in stylized market models where a single shock to the value
occurs, the price converges rapidly to an equilibrium price,
without expected loss; further, the markets need not have bounded payoffs.
The drawback of these information based market makers has thus far
been that after quick convergence following an initial market shock to
the true value, the convergence after a subsequent market shock 
is slow, because the market maker gets ``overconfident'' after initial
convergence.\footnote{%
If the subsequent shock occurs at time $t$, the time to converge to
the new equilibrium value is exponential in $t$.
}  

The behavior of the LMSR and information based market makers are
discussed further in Section~\ref{section:mm}.  One usually compares
market makers either theoretically (eg. \cite{chen2007utility}) or by
using simulation in some stylized model of trading
(eg. \cite{Das08}). All the properties discussed above are evident
within such stylized models. There has been little systematic
exploration of the performance of various market makers in real
settings, where trader behavior is unpredictable.  In this paper we
present a new information based market maker which is able to adapt to
multiple market shocks, and evaluate it using a novel experimental
design for comparing market microstructures in live trading
experiments with human traders.

\paragraph{Contributions}
We introduce a new information based market maker 
which builds from the zero profit market maker in 
\cite{Das05a,DasMag08} -- we call this market maker BMM (for
Bayesian Market Maker). The main innovation is the
ability to adapt to multiple market shocks.
BMM 
provides liquidity by adapting
its spread based on its level of uncertainty about the true value. This
allows
it to achieve small spreads in equilibrium-like states,
while remaining adaptive to shocks
by increasing its spread when the
information content in the population changes.

To adapt rapidly to multiple shocks, BMM can increase its 
uncertainty parameter
exponentially quickly. It does so
 by constantly comparing the probability of observed trader 
behavior over the recent history under its current uncertainty with that under 
the increased uncertainty.
This ability to adapt to shocks has a drawback. There is always a
random 
chance that recent history will cause BMM to increase
its uncertainty level, and hence increase spreads, even when there
is no actual shock to the market. This means that convergence to the true
value during an actual equilibrium will occasionally get interrupted
by chance fluctuations; however these chance fluctuations are minor
compared to the (typical case) 
fluctuations of the LMSR market maker. Nevertheless, these 
fluctuations represent a real tradeoff between the tightness of convergence 
during an equilibrium and the ability to adapt quickly during a 
market shock. By tuning the extent of recent history used in 
determining BMM's uncertainty level, the market designer can quantitatively
control this tradeoff.

In simulations, as well as in real trading,
BMM can provide substantial benefits compared
to the Hanson LMSR market maker in many situations, with the caveat
that it does not provide a similar guarantee on maximum loss -- thus,
there are situations where its performance (with respect to loss incurred) 
can be worse; such situations are extremely adversarial, however.
As with the $b$ parameter in Hanson's LMSR, the dynamics of the uncertainty
level in BMM 
controls the tradeoff between
how adaptive the market maker is to changes in market conditions and
its potential loss. However, the nature of this dependence is different.

Our second main contribution is to develop an experimental paradigm
for comparing market microstructures. In particular, we apply this to
compare BMM and Hanson's LMSR market maker 
in a live trading setting. 
Two challenges one faces with live
trading are: 1) the same group of traders cannot 
be used first in an experiment with one market maker, and then in a
second identical experiment with a second market maker. This is because 
traders get primed, and even if the experiments
are identical, the results are incomparable;
2) the same experiment cannot be run on two separate groups of traders
with a different market maker in each group, because the high variability
in human traders results in a very
high variance due to the small size of any such experiments with human
traders.
The ideal experiment simultaneously tests both market makers on the
same population of traders in a \emph{symmetric way}.

We present a novel experimental design which can capture many 
aspects of the way information is continuously 
revealed to traders, in addition to 
allowing for market shocks with and without visually perceptible
cues. Our design is based on a graphical 2-dimensional random
walk which simulates the classic \emph{Gambler's Ruin} problem and
allows us to symmetrically compare different market structures.  
We demonstrate the design with several experiments.
Our experimental results confirm our results from simulation, namely that
even in a real setting, with unpredictable human traders, BMM outperforms
Hanson's LMSR.

In summary, this paper introduces a new market making algorithm, BMM,
that has great
potential for prediction markets. It offers two compelling
advantages over the Hanson LMSR market maker. First, at any point in
time, it provides a meaningful and useful distribution for the
probability of an event occurring. Second, in expectation, it makes
less loss while providing an equally liquid market (however, the
corresponding disadvantage is that it is not loss-bounded). The
development of this algorithm also provides strong evidence that there
is a fundamental tradeoff between the reactivity of a market maker to
changes in market conditions and the amount of loss it can be expected
to suffer. 
Despite the complex underlying mathematics, the ultimate implementation of
BMM is 
simple, computationally efficient and can be succinctly
described. 

We demonstrate the benefits of BMM in several experiments with human participants. In doing so, we also introduce a new family of experiments that can be used to compare market makers (or, indeed, entire market microstructures) in a fair, symmetric manner. These experiments can be extended far beyond the present work.

\paragraph{Paper Outline}
We continue next with a brief introduction to inventory and information based market making.  We then discuss our new information based market maker BMM which can adapt to multiple market shocks. We compare the market makers in both simulation and live trading, and conclude with some open questions and interesting avenues for future work.

\section{Market Making}
\label{section:mm}
The key challenge in most prediction markets is inducing sufficient
liquidity. 
How can one incentivize participants
with good information to trade? Without uninformed
traders to make profit off of,
informed traders will not trade (the 
No-Trade theorem of Milgrom and Stokey
\cite{milgrom1982information}). 
A means of creating ``uninformed'' (or less informed) trades and
thereby providing liquidity in modern online prediction markets is
through automated market making algorithms \cite{Han07,PenSam07}. A
market maker is an intermediary willing to take the other side of
every trade, buying (resp. selling) when someone wants to sell (resp.
buy); the market maker sets the prices, which will affect whether the
trade will actually execute or not. We consider a pure
dealership market, where a market maker takes one side of \emph{every
  trade}. An apt comparison could be the foreign exchange desks at
airports, typically a monopoly for Travelex, who get to set bid and
ask prices for foreign currency transactions. This model of the market
allows us to compare market makers in a fair and precise manner, but
in the future it will be important to consider integrating market
makers with limit order books (which poses more of a challenge for
evaluation than design).

\subsection{Inventory-Based Market Making}
Hanson describes a market maker  for combinatorial
prediction markets \cite{Han07}, which we briefly
review here in the context of a single market.
Hanson's technique adapts the idea of a scoring rule to a
prediction market setting. While many different scoring rules are
possible, Pennock reports that in practice the logarithmic scoring
rule is the most useful.\footnote{%
\url{http://blog.oddhead.com/2006/10/30/implementing-hansons-market-maker/}} 
The market maker will take the opposite side of any order at a price 
specified by the market maker. This price depends on a parameter 
\math{b} and the market maker's current inventory \math{q_t}, where
\math{t} indexes the arrival of trade requests; the 
inventory starts at zero, \math{q_0=0}, which corresponds to an initial
price of \math{0.5}.
The market maker sets prices so as to guarantee bounded loss, no matter
what the true liquidation value is.

Specifically, the spot price is given by
\mand{
\rho(q_t) =
\frac{e^{q_t/b}}{1+ e^{q_t/b}}.
}
At time \math{t+1}, if a trade arrives for quantity \math{Q}, the 
cost of the trade (to the trader) is given by 
\mand{
C(Q;q_t)=\int_{q_t}^{q_t+Q}ds\ \rho(s)=b\ln(1+e^{(q_t+Q)/b})-b\ln(1+e^{q_t/b}).
}
The volume weighted average price is \math{|C(Q;q_t)|/q}, and it corresponds
to the trader accruing a position of size \math{Q} using infinitesimal 
increments, paying the prevailing spot price at each increment. If the
trader accepts the trade at this average price, then the market maker 
updates its state to \math{q_{t+1}=q_t+Q}. Since the 
starting inventory \math{q_0=0}, it is easy to verify that the 
maximum loss incurred by the market maker is 
\math{\lim_{Q\rightarrow\infty}Q-C(Q;0)=b\ln 2}.

The parameter $b$ is the only free parameter in the LMSR
market maker; not only does it bound the loss of the
market maker, but it also controls how adaptive the market maker is.
If \math{b} is small, the market maker is very adaptive, taking on small 
loss;
 \math{b} also controls liquidity in the market. An adaptive market maker leads
to large bid-ask spreads, implying less liquidity.

It is known that a market mediated by the LMSR market maker can yield
a rational expectations equilibrium if traders incorporate information
from prices into their beliefs in a rational manner. However, consider
what happens in a case where a large trading population continues to
maintain somewhat different beliefs, and some traders regularly come
in and trade some typical trade size \math{Q}. The bid-ask spread
\math{\delta(Q)} for quantity \math{Q}, 
given the
current inventory \math{q_t}, is the difference between the average 
price paid for buying \math{Q} shares versus selling \math{Q} shares;
\mand{
\delta(Q)=
\frac{b}{Q}\ln\left(\frac{\cosh q_t/b+\cosh Q/b}{2\cosh^2q_t/2b}\right).
}
At market inception (\math{q_t=0}), the spread is decreasing in \math{b},
so 
higher $b$ means more liquidity (in general the relationship between 
liquidity and \math{b} is not monotonic).
Suppose the equilibrium price corresponds to an inventory \math{q_{\text{eq}}};
if typical trade sizes are \math{Q}, then the spot price fluctuations 
around this equilibrium have magnitude
\mand{
\frac{\sinh(Q/b)}{\cosh(q_{\text{eq}}/b)+\cosh(Q/b)}.
}
These fluctuations are asymmetric about the equilibrium and persist, making
it hard to extract a quantitative probability estimate. The choice
of \math{b} is an important open problem; smaller \math{b} guarantees smaller
loss, but  a less liquid market with higher fluctuations around 
the equilibrium. Figure \ref{fig:motiv}
compares the LMSR market maker to
an information based market maker 
(which we discuss next) in a highly stylized model. As we see, the 
LMSR market maker is adaptive but non-convergent;
the information 
based market maker ZP \cite{Das05a,DasMag08}
is convergent, but only slowly adaptive; a slowly adaptive
market maker will incur large loss. 
One of our goals is to improve ZP to have better equilibrium convergence (less fluctuation) than the LMSR, while still being able to adapt quickly.

\subsection{Information-Based Market Making}

\begin{figure}[!h]
\begin{minipage}[c]{0.48\textwidth}
\begin{center}
\includegraphics*[width=2in]{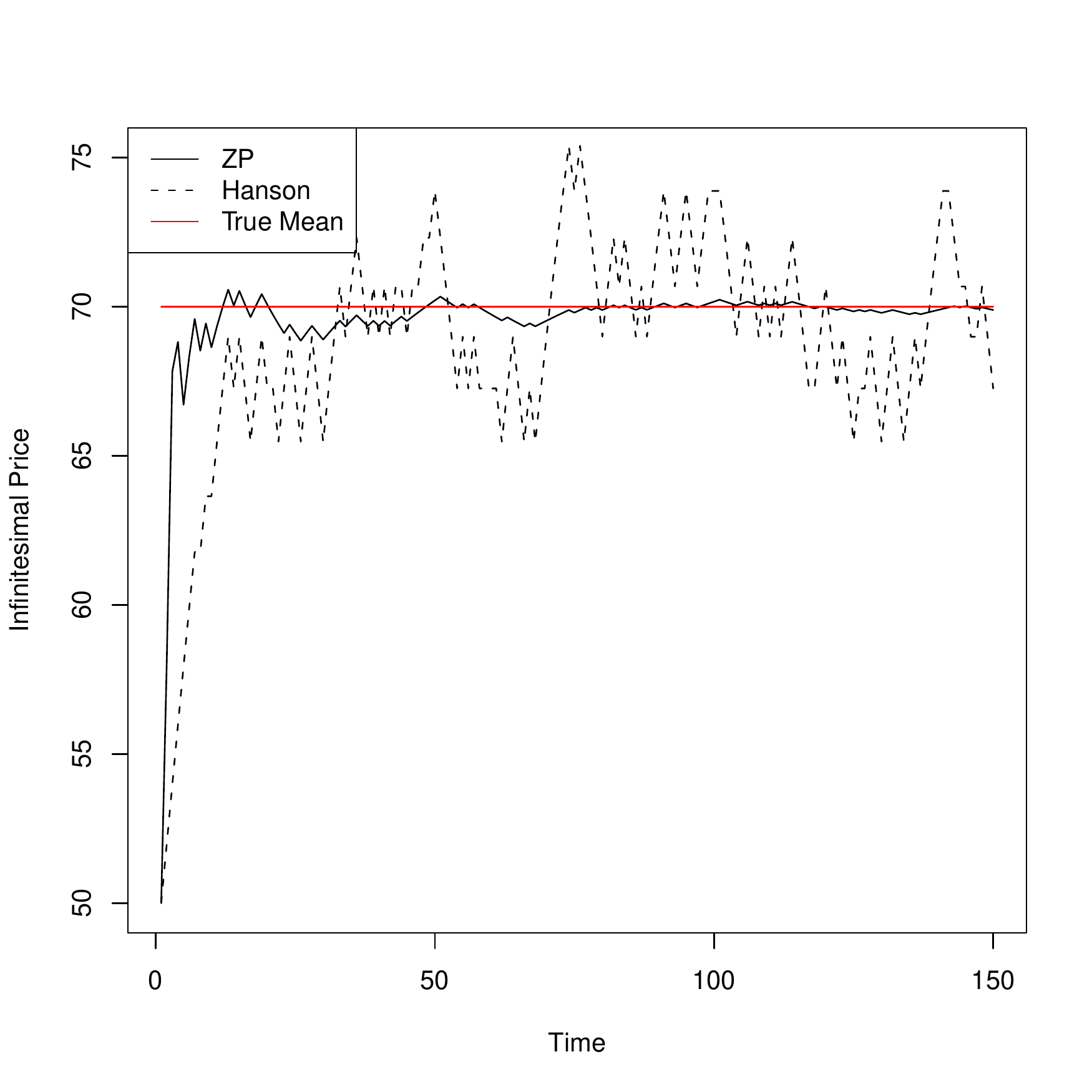}\\
(a) Equilibrium behavior in a stable market.
\end{center}
\end{minipage}
\begin{minipage}[c]{0.48\textwidth}
\begin{center}
\includegraphics*[width=2in]{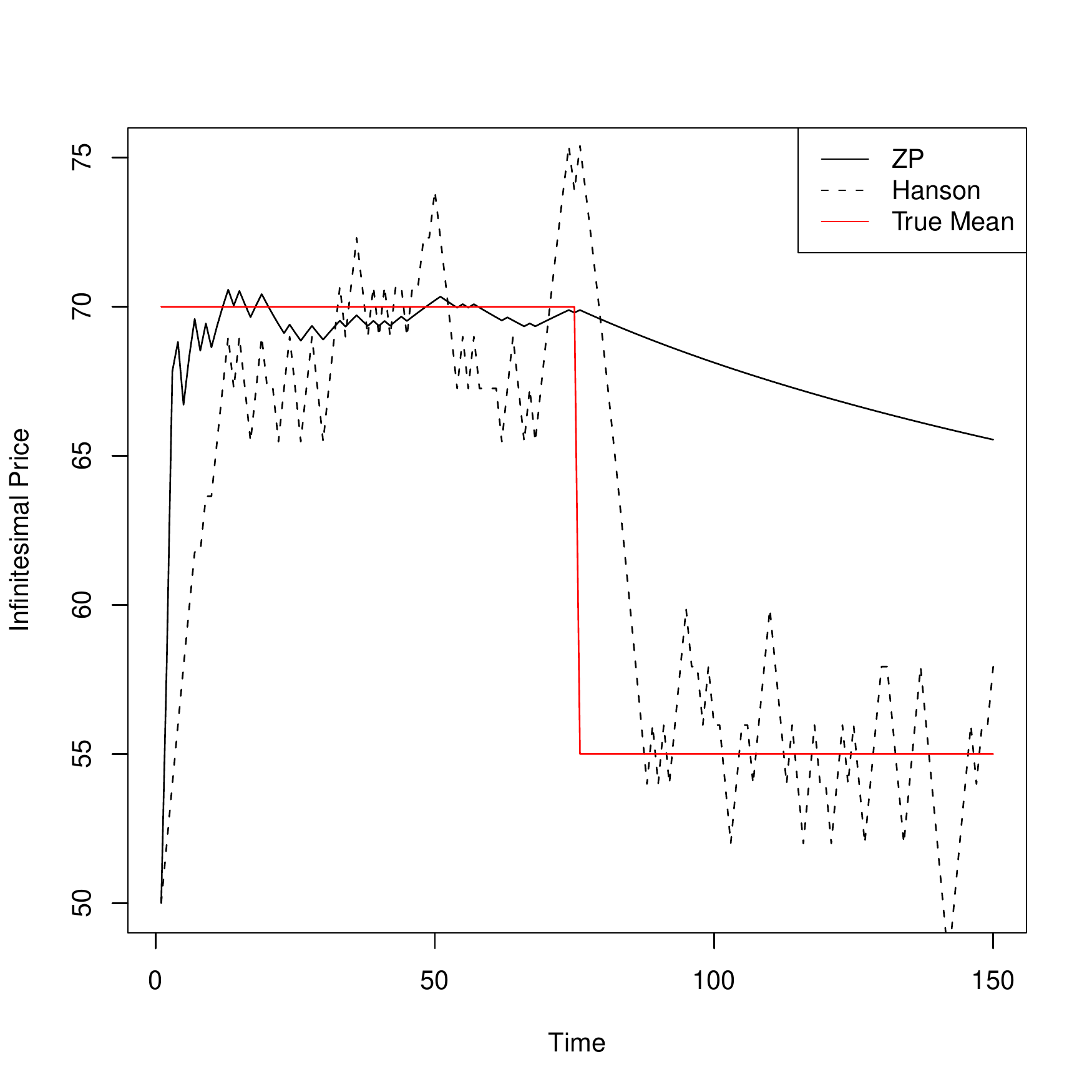}\\
(b) Adapting to a market shock.
\end{center}
\end{minipage}
\caption{Behavior of the Hanson LMSR market maker and the Bayesian
information based market maker ZP \cite{Das05a,DasMag08},
for an idealized setting. In both experiments, trades are a fixed size, and
traders receive a value signal on the basis of which they trade.
In (a) the trader value signals are distributed about a stable mean; in 
(b), there is a jump in the mean.
For the LMSR market maker, the parameter \math{b} was set so that the
average spread of the two market makers matched.
In (a), we see that both market makers rapidly arrive at the true 
value, but LMSR fails to converge, whereas the information based
market maker converges.
From (b) we see that the LMSR market maker 
is very adaptive to a market shock
whereas the Bayesian information based market maker
can only adapt very slowly because it becomes overconfident. 
These figures are based on simulations where traders have beliefs 
drawn from a Gaussian distribution (the mean of which jumps 
in the second case), and trade in fixed size blocks.}
\label{fig:motiv}
\end{figure}

Though the LMSR market maker is loss-bounded, being
purely inventory-based, it is an
extremely uninformed trader; it typically will
substantially subsidize the market, taking on large loss. 
Alternative market making schemes necessarily incur more risk. In
Hanson's words, ``a computer program with less than human intelligence
that attempts to make markets runs the risk of being out-smarted by
human traders'' \cite{hanson2009on}. This is because a market maker who makes
offers to buy and sell any security runs the risk of losing out to
either better informed or smarter traders. At the same time, ``smart''
market making algorithms may be able to exploit human
trader errors or overconfidence. 
Thus, it might be possible to provide liquidity without
substantial loss. 

The market maker which we present is information based, 
and builds on the 
zero profit market maker 
in \cite{Das05a,DasMag08}. We first briefly describe this
model of market making, postponing the details to 
Section~\ref{section:algo}.
We start from the canonical Glosten-Milgrom model
of price-setting under asymmetric information \cite{GloMil85}.
At time \math{t}, the market maker has some belief (prior
probability density) about the value of the security
\math{p_t(v)} (which is the mean of the distribution on trader valuations). 
We assume $p_0(v)$  is correct, so the realized value \math{V \sim p_t(v)}.
 An arriving trader gets a signal 
\math{s} drawn from a distribution whose expectation is \math{V};
the variance of \math{s} measures the uncertainty in the trader's signal
(or \emph{information set}). The market maker's only information is its prior
belief on \math{V}. Hence the information available to market maker and
trader are different, and this information asymmetry can be measured by the
information disadvantage of the market maker, the ratio of the variance in 
the market maker's prior belief and the trader's uncertainty. This information
disadvantage plays an important role in the market maker's actions.

Given this initial setting, the market maker must set a bid and ask price, and
the trader will trade accordingly: if 
\math{s<\text{bid}}, the trader will sell, and if
\math{s>\text{ask}}, the trader will buy.
In a competitive setting, the market maker will set prices so as to receive
zero expected profit. To do this, the market maker solves two non-linear
fixed point equations,
\eqan{
\text{ask}&=&E_{p_t(v)}[v|s>\text{ask}],\\
\text{bid}&=&E_{p_t(v)}[v|s<\text{bid}].
}
The sequence of papers \cite{Das05a,Das08,DasMag08} extend this model to the
sequential setting with a Bayesian learning market maker. After setting
prices, the market maker can now observe what the trader does (buy, sell or no
trade). This gives the market maker information regarding the trader's signal
\math{s}, and hence information regarding the realization \math{V}. Thus, the 
market maker can update its prior beliefs \math{p_t(v)} to 
\math{p_{t+1}(v)} to incorporate this new information. The market maker
is now ready for the next trader.

The learning market maker in the sequential model 
is composed of two related parts. The first maintains the
belief distribution on the value of the market, \math{p_t(v)};
the second sets prices to achieve some goal, for example zero expected
profit. 
From the reinforcement learning perspective, 
bid and ask prices serve as actions, and agents' decisions 
to buy or sell at those prices provide observations that allow
 the market maker to update its beliefs. As in most reinforcement
 learning problems, the actions (prices) serve the dual role of
1)  eliciting
information (setting the bid-ask spread too high will lead
 to a lack of trading, yielding little information 
about the trading public's beliefs) and 2) generating reward.

Das and Magdon-Ismail \cite{DasMag08} present efficient approximate
algorithms for performing these updates for zero profit (ZP) as well
as profit maximizing monopolist market makers.  In the specific algorithm
considered, the trader signals are drawn from a Gaussian distribution,
and the initial market maker belief is also Gaussian.  They show that
convergence occurs quickly (the market maker's uncertainty drops to
zero), starting from a state with high information disadvantage.  This
is an advantage over the LMSR market maker which cannot converge in
equilibrium unless all trader beliefs also converge. When the market
maker's initial belief is nearly correct, these algorithms deliver
zero or near maximum expected profit, another advantage over the LMSR
market maker. However, in the event of a market shock, the convergence
to the new market value is exponentially slow 
-- i.e. the market maker is not very adaptive. Figure \ref{fig:motiv}
compares ZP to the LMSR market maker in the stylized model that
corresponds to the assumptions used to derive the ZP equations,
illustrating the pros as well as the cons. Clearly what is needed is a
market maker with better convergence properties than LMSR but better
adaptability than ZP.

\section{Market Microstructure}
\label{section:model}
We consider a prediction market with a single binary outcome stock
that trades between 0 and 100. Presumably, if the event occurs it pays
off 100, and if not, it pays off 0. However, this can also be thought
of as a stock with a liquidating dividend between 0 and 100. At any
point in time, an arriving trader sees the history of trading in the
stock, and the ``current price,'' which can be thought of as either
the infinitesimal price, the market maker's mean belief about the
probability of the event occurring, or the middle of the bid-ask
spread. The trader then chooses a quantity that she wants to buy or
sell. The market maker observes the quantity demanded, and sets a
price based on this quantity. The trader is informed of this price and
can then choose whether or not to execute the trade at that price.

Our market is structured as a pure dealer market, with the market
maker as the only price setter. Only a single (infinitesimal)
spot price is seen by arriving traders. This is a natural
formulation for the Hanson LMSR market maker, because the actual price
is a continuous function of the quantity demanded. However, it also
serves an important theoretical role for BMM. If an arriving trader
saw simultaneous bid and ask prices and had a valuation between the
two, she would not initiate any kind of trade, leaving the market
maker with no information about her valuation. Especially in times of
high uncertainty, this lost information can be a major source of lost
liquidity. When an arriving trader sees only a single infinitesimal
price, she is inclined to ``test'' the market by placing an order on
one side or the other. Then, even if she does not execute the trade,
the market maker can glean valuable information about where her
valuation lay (if she places an order but then does not execute, the
market maker can infer that her valuation lay between the
infinitesimal price and the quoted price).

\section{The BMM Algorithm}
\label{section:algo}
\providecommand\remove[1]{}

BMM is based on the zero-profit market maker (ZP)
mentioned in Section~\ref{section:mm} and described in 
detail in \cite{DasMag08}, with two main innovations:
1) the ability to deal with trade sizes; and, most importantly,
2) the ability to adapt quickly to market shocks.
A trader arrives, observes the spot price \math{p_t} and
requests a trade for quantity \math{Q} in a direction \math{x_t=\pm1}. 
\math{x_t=+1} means the trader would like to buy. For concreteness, we will
assume that \math{x_t=+1}; however, the process is completely symmetric.
The market maker performs 3 tasks.
\begin{enumerate}[i.]
\item Provides a VWAP quote for Q shares;
\item Updates its state depending on whether the trade is 
accepted or canceled;
\item Maintains a validity measure for its current beliefs, which is 
crucial to being able to adapt to market shocks.
\end{enumerate}
We briefly summarize ZP described in \cite{DasMag08} first.
The market maker's state \math{p_t(v)}
is characterized by a Gaussian
belief  for the value of the market $V$: \math{p_t(v)=N(\mu_t,\sigma_t^2)}.
The trader signal is assumed to be normally distributed around \math{V}, so 
\math{s\sim N(V,\sigma_\epsilon^2)}.
The main relevant parameter
(see \cite{DasMag08}) is the information disadvantage of the market
maker,
$\rho_t = \sigma_t / \sigma_{\epsilon}$, the ratio of the uncertainties of the 
market maker and trader.
A universal ``Q-function'', \math{Q(\rho)} (see \cite{DasMag08})
plays an important role in quoting prices.
Specifically,
the spot price is just the market makers mean belief, 
\math{p_t=\mu_t}, and the ask price is 
\mand{\text{ask}=\mu_t+\sigma_\epsilon Q(\rho_t)\sqrt{1+\rho_t^2}.}
This ask gives zero expected
profit conditioned on the trade going through;
this quoted price does not take quantities into account. Described in 
\cite{DasMag08} is a range based  update procedure for the 
market maker: if a trader's realized signal is known to lie in the range
\math{z^-<s<z^+}, then the market maker 
updates its Gaussian belief to:
\eqan{
\mu_{t+1} &=& \mu_{t} + \sigma_{t} . \frac{B}{A},\\
\sigma_{t+1}^2 &=& \sigma_{t}^2 \bigg( 1 - \frac{AC + B^2}{A^2} \bigg),
}
where \math{A,B,C} are functions of 
\math{z^-,z^+,\mu_t,\rho_t,\sigma_\epsilon}, the details of which are given
in \cite{DasMag08}. This range based update is used when the trader
takes an action (accept or cancel the trade). So, for example, if the 
trader accepts a trade, then \math{s\in[\text{ask},\infty)}, and so
\math{z^-=\text{ask}} and \math{z^+=\infty}. If the trader cancels upon seeing the quoted price, \math{z^-= \mu_t} and \math{z^+=\text{ask}}.

\subsection{Quoting a Price for Q Shares}
ZP can only quote a price for a fixed trade size. To be practical, the
algorithm needs to quote a price for an arbitrary number of shares. The 
spot price is \math{p_t=\mu_t}, and assume a trader wants to buy
\math{Q} shares.
We implement a heuristic of treating this order as independent orders
of a fixed size \math{\alpha}. There are thus \math{\ceil{Q/\alpha}}
independent orders; the sizes \math{\alpha_1,\alpha_2,\ldots,\alpha_k}
are all \math{\alpha}, except possibly the last one.

The market maker starts in state \math{\mu_1=\mu_t}, 
\math{\sigma_1^2=\sigma_t^2}, and 
imagines the arrival of these \math{k} mini-orders in 
sequence;
for each mini-order arrival, the market maker quotes the ZP price as in 
\cite{DasMag08}; each mini-trade is accepted; the market maker then
updates his belief and receives the next mini-trade. Specifically, consider
mini-trade \math{i}, with market maker belief \math{\mu_i,\sigma_i^2}.
The price quoted is 
\mand{
p_i=\mu_i+\sigma_\epsilon Q(\rho_i)\sqrt{1+\rho_i^2};
}
the trade is accepted, so the market maker updates his belief with 
\math{z^-=p_i;\ z^+=\infty}:
\mand{
\mu_{i+1}=\mu_i+\sigma_i\frac{B}{A},\qquad
\sigma_{i+1}^2=\sigma_{i}^2 \bigg( 1 - \frac{AC + B^2}{A^2} \bigg);
}
the market maker now processes the next mini-order in the sequence until all
the mini-orders are processed. Note that these mini-orders are not real,
they just describe the process going on in the market maker algorithm. Thus,
\math{\alpha_1,\ldots,\alpha_k} shares, with
\math{\sum \alpha_i=Q},
get (fictitiously) executed at the
prices \math{p_1,\ldots,p_k}. The price quoted to the trader for \math{Q} 
shares is the VWAP for this fictitious sequence of executions:
\mand{
\text{ask}=p(Q)=\frac{1}{Q}\sum_{i=1}^k\alpha_i p_i.
}
\paragraph{Belief Update}
Since the trader asked to buy, we know that \math{s\ge p_t}. The trader is
quoted a price \math{p(Q)}, and so based on the traders action, the market 
maker can update his beliefs to \math{\mu_{t+1},\sigma_{t+1}^2} using the 
range update:
\mand{
\text{trade}
\begin{cases}
\text{accepted}&z^-=p(Q);\ z^+=\infty;\\
\text{canceled}&z^-=p_t;\ z^+=p(Q).
\end{cases}
}
We described a buy order, but a sell is entirely symmetric.

\subsection{Adapting to jumps}

The original ZP algorithm leads to constantly decreasing variance
 of the market maker's belief. After a number of trades have been 
processed the variance and therefore the spreads are significantly reduced. 
While this increases liquidity and encourages further trading towards the 
true market valuation, it is also the root of 
the market makers inability to adapt to multiple market shocks.
In fact, the magnitude of each mean belief update is proportional
 to the variance of the market-maker's belief, 
large jumps in the true underlying value coupled with a 
small belief variance, lead to very small update values, 
and the algorithm is exponentially slow in adapting to a jump.

After a jump, the sequence of trades will be ``one-sided'', and 
hence inconsistent with a market makers belief of the old valuation 
coupled with a highly confident low belief variance. The simple solution
to this is to allow the market maker to become less confident as he
see a sequence of extremely one sided trades, i.e. an inconsistent
sequence
of trades. To accomplish this, we define a consistency index
\math{C(\text{history})}, which measures exactly how likely the recent
history of trades observed under the current uncertainty level is, as 
compared to a higher uncertainty.
An intuitive solution is 
to increase the market maker's belief variance during periods of 
inconsistency. 

Specifically,
BMM keeps track of a fixed window of previous trades (including
canceled trades), 
along with the $z^{+}$ and $z^{-}$ values that are inferred from those trades.
Then, at 
a particular time step, the probability of a sequence of trades
over a window of size $W$, can be computed as:
\[ L (\mu, \sigma) = \int^{\infty}_{-\infty}  \! N \big(v, \mu, \sigma
\big) .  \prod_{i=1}^{s} \bigg(\Phi \big(z^{+}_{i}, v,
\sigma_{\epsilon}\big) - \Phi \big(z^{-}_{i}, v,
\sigma_{\epsilon}\big)\bigg)  \, dv \]
The intuitive solution is to compare this
probability against a fixed threshold; if the probability is too small, we are
in an inconsistent regime, and so we increase the market maker's uncertainty 
level (increase the variance).
However, this solution is problematic because the threshold is 
highly sensitive to the choice of window size 
and particular features of the trade sequence. 
Instead, we make a relative comparison with 
 the same probability computed at twice the uncertainty.
We thus define our consistency index
\mand{
C(\text{history})=L(\mu_t,2\sigma_t)-L(\mu_t,\sigma_t);
}
If \math{C>0}, we increase \math{\sigma_t}, specifically
\math{\sigma_{t+1}=2\sigma_t}.
The choice to double the variance is arbitrary, and any multiplier greater
than 1 would do.
Though we have only tried the multiplier \math{2}, we expect that
since this is a relative measure of consistency, the results would be 
robust to the choice of multiplier, unlike with the use of a fixed 
threshold.

This algorithm takes advantage of the fact that more ``even'' sequences 
of trades are more likely when the variance is lower, 
while sequences that are heavily biased in one direction or 
the other become more likely with higher variance. 
The key parameter for this algorithm is the window size $W$, 
which controls the balance between how stable the market maker 
is at equilibrium and how fast it can adapt to changes. The window size 
$W$ also now becomes the dominant factor in measures 
like average spread, so that the particular value of $\sigma_{\epsilon}$ 
becomes unimportant.

\section{Simulation Experiments}
\label{section:sim}
In order to test the market making algorithms and elicit their general properties, we conducted extensive simulation experiments before deploying them in situations with live human trading. The goals were to (1) ensure the adaptive capabilities of BMM (2) compare BMM and Hanson's LMSR MM on the basis of profit/loss, average spreads, and price discovery, and (3) calibrate parameters so that the live trading tests could be done with market makers that were similar in average spreads.

The simulation environment is structured as follows. Each trading simulation consists of 200 discrete time steps. There is an underlying ``true value'' process. The initial true value is drawn from a Gaussian distribution with mean 50 and standard deviation 12 (in general, all values are truncated at 0 and 100 whenever that may be an issue). Then, at every time step, there is a probability $p_j$ that the true value jumps. We consider two different types of jumps. In the first type, which is more realistic, the amount of the jump is drawn from a Gaussian distribution with mean 0 and variance $\sigma_j^2$. In the second type, which is meant to simulate a very problematic case for an information based market maker, the new value is itself drawn uniformly at random between 0 and 100. At any point in time, an arriving trader receives a valuation $w_t$ drawn from a Gaussian distribution with mean equal to the true value at that time, and variance $\sigma_{\epsilon}^2$.  If $w_t$ exceeds the current infinitesimal price, the trader initiates a buy order, and if it is less the trader initiates a sell order. The quantity is drawn at random from an exponential distribution with rate parameter $\alpha_q$.\footnote{%
This random quantity model is frequently used in models of zero-intelligence trading and models from the econophysics literature (e.g. \cite{FarPat05}).
} %
In our experiments, we set $p_j = 0.01$, $\sigma_j = 5$, and
$\sigma_{\epsilon} = 5$. $\alpha_q$ is set to $0.05$ so that the mean
trade size is 20.  The $b$ parameter for the  LMSR market maker was
set to 125 and the window size parameter for BMM was set to 5. These
choices of the MM parameters were in order to make the average spread
approximately equal in the Gaussian jumps case, and were then used
again for the initial live trading experiments described in the next
section.

\begin{figure*}[t]
\begin{center}
\begin{tabular}{m{0.32\textwidth}m{0.32\textwidth}m{0.32\textwidth}}
\includegraphics*[width=0.29\textwidth]{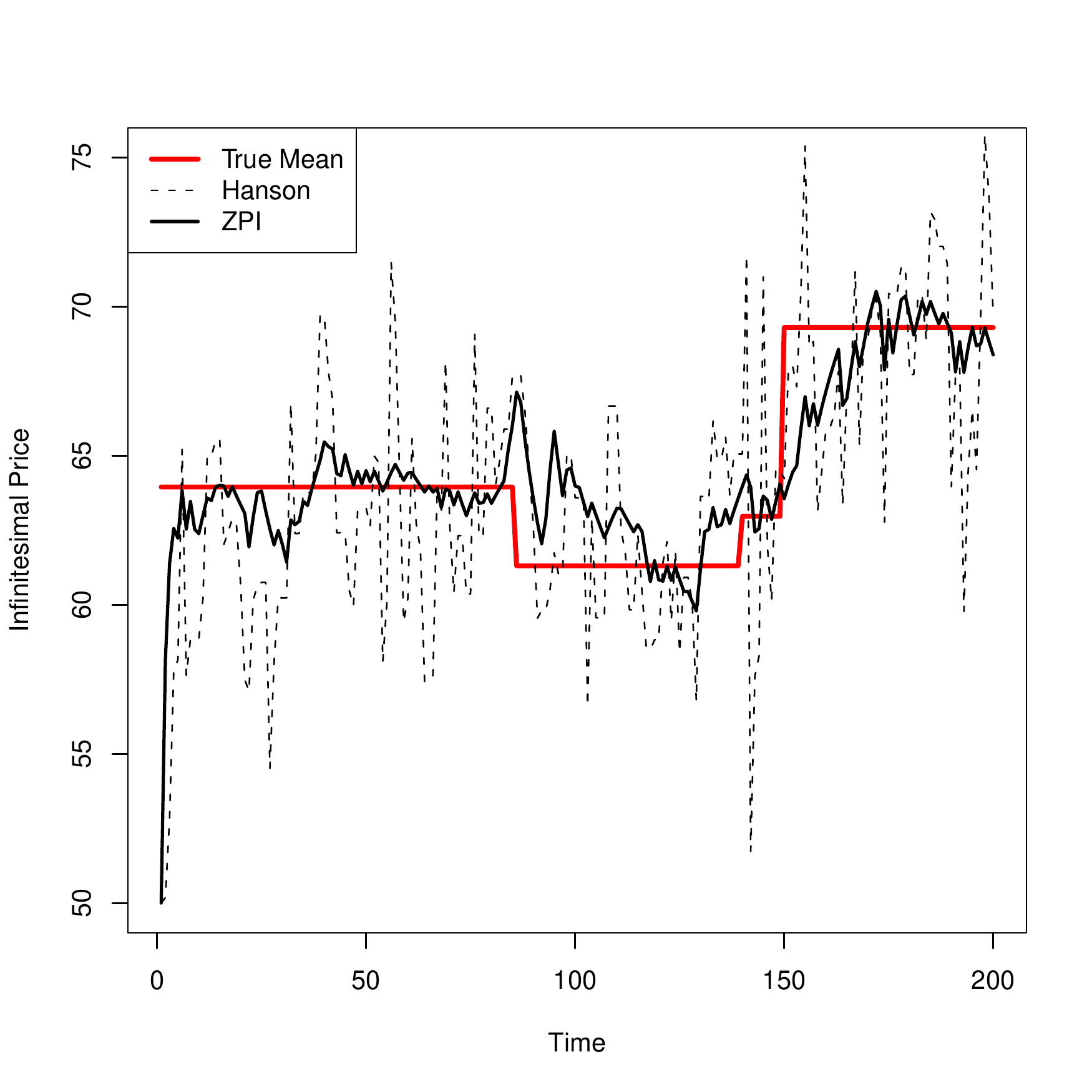}&
\includegraphics*[width=0.29\textwidth]{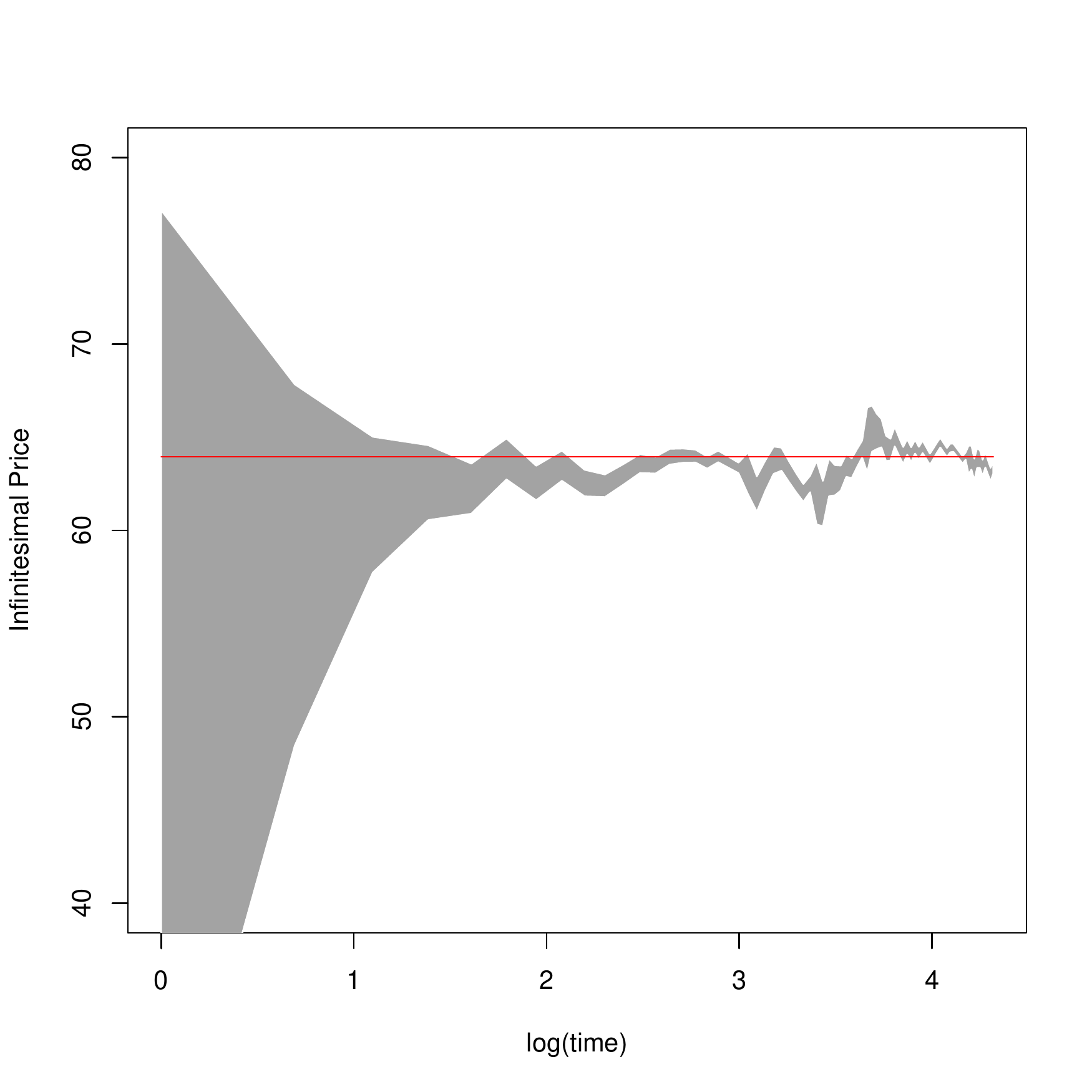}&
\includegraphics*[width=0.29\textwidth]{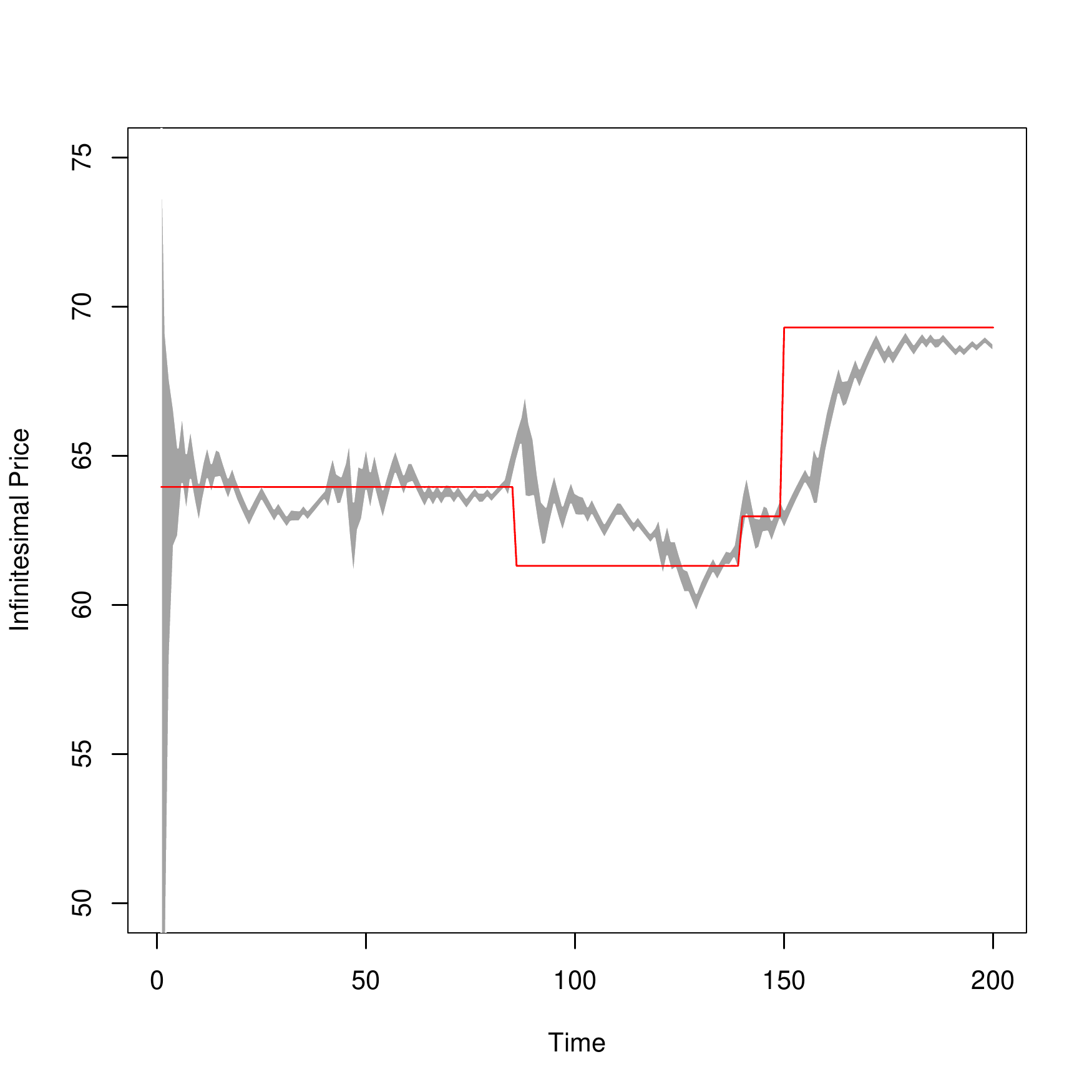}\\
Spot price versus true market value.
&
Initial convergence of MM's beliefs (spread) shown by the width of the
gray region (log scale). 
&
Spread convergence for window size 10.
\end{tabular}
\caption{Behaviors of the market makers in an example simulation. The first two figures show BMM with window size 5, and the third shows window size 10. BMM is clearly capable of adapting rapidly to changing behavior in the trading population, but at the same time shows less jumpy and unstable behavior at equilibrium than the LMSR MM. This behavior is explained by the spread (which is a function only of the variance of the MMs belief). The spread starts off high, and increases around times of uncertainty, allowing the mean to move more quickly. At the same time, this can create occasional instabilities even when the underlying population mean has not jumped (note some of the periods of increasing spread when the true value is stable).
\label{fig:anecdote}
}
\end{center}
\end{figure*}

Figure \ref{fig:anecdote} gives some intuition into the behavior of BMM as compared with LMSR. This is for a single experiment, and shows that BMM can adapt rapidly to changing valuations in the trading population, while at the same time settling into periods of low spreads and stable behavior at equilibrium. The typical behavior is to start off with a high variance (and hence high spread), and then quickly converge to a low variance regime. When a jump in the population belief occurs, the market maker can quickly pick up on that fact using the algorithm described previously, because the sequence of trades it sees is usually heavily biased in one direction, which would be more likely to occur if the market maker's beliefs had a higher variance (in contrast, series of trades that are more balanced are more likely to occur in a model with lower variance, since the probability mass is more concentrated in the ``likely'' region). Because of the adaptivity, in a long stable period there will be times when the variance (and spread) will increase even though no true change has occurred. This becomes more likely as the variance gets lower. 

It is important to point out that this behavior is general. For about
the same average spread, BMM can in general achieve better market
properties in terms of stability at equilibrium as well as profit. In
this particular simulation, the average quoted (half) spread for BMM
was $1.23$ and its profit was $3080.29$. The average quoted (half)
spread for the LMSR MM was $1.91$ and its profit was $-956.27$.  Table
\ref{table:simresults} demonstrates this fact more generally by
showing results from 1000 simulations. In addition to the average
profit  and spread, this table also reports the root mean square
deviation of the infinitesimal price from the true value (population
mean) at any given point in time (a measure of price discovery), and
the single worst loss suffered by the market makers in 1000
simulations (in both cases the single worst loss suffered by the LMSR
market maker is close to the theoretical bound of $8664.34$. BMM
performs better on average. However, it is worth noting that, as the
probability of a jump goes up, especially in the case where new
valuations are drawn uniformly at random, the loss suffered by BMM
increases, so it may not be the best choice for highly unstable
environments.  

\begin{table}
\begin{center}
\begin{tabular}{l|ll|ll|}
 & \multicolumn{2}{c|}{Gaussian Shocks} & \multicolumn{2}{c|}{Uniform Shocks} 
\\
 & BMM & LMSR & BMM & LMSR\\ \hline
Profit & $2081.35$ & $-2457.30$ & $603.40$ & $-1897.98$\\ 
Max Loss & $9479.82$ & $8662.32$ & $50183.77$ & $8384.42$ \\
Spread & $1.42$ & $1.35$ & $1.79$ & $1.40$ \\
RMSD & $2.92$ & $5.38$ & $8.78$ & $10.79$ \\
\end{tabular}
\end{center}
\caption{Performance of BMM and the LMSR MM in simulated trading
\label{table:simresults}
}
\end{table}

\section{Live Trading}
\label{section:expt}
\begin{figure}
\includegraphics*[width=\textwidth]{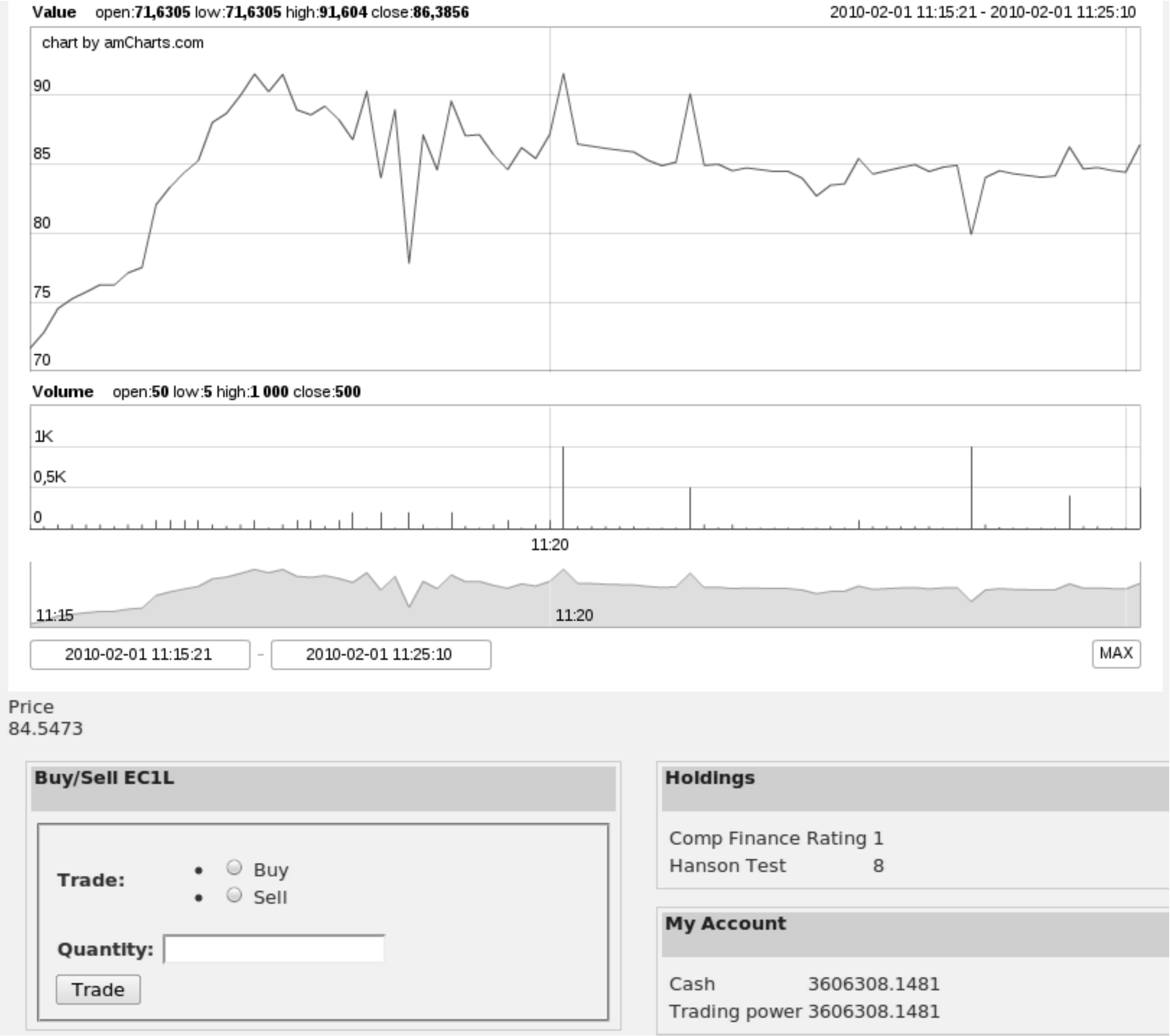}
\caption{Trading interface}
\label{fig:interface}
\end{figure}
We now
present an experimental design for comparing market makers in a live
trading setting with human subjects. 
Human subject experiments by their very 
nature use small samples; further, human subjects are diverse and
very rapid 
learners, whose attention cannot reliably be maintained for 
extended time periods.
This poses several challenges to live trading experiments when
trying to  compare market
makers.
\begin{enumerate}[(i)]
\item 
Two comparably sized groups can display vastly different behaviors
due to inherent diversity in backgrounds, skill sets and tendencies among
human subjects.
\item
Human subjects, being natural learners, build biases very quickly. So, 
for example, if you run an experiment for the first time with a market 
value of (say) 0.7, traders may take some time to become accustomed
to the trading task. If you run exactly the same experiment again, it is
possible that the second time around, the traders will display more 
intelligent behavior, with perhaps even a bias that the value is around
0.7, having ``generalized'' from the previous experiment.
\end{enumerate}
The implications are that to get useful results, the live trading
experiment should use the same group of traders simultaneously to compare
a pair of market makers. Further, the market makers should be 
compared in a completely symmetric way, using an intuitive interface.

We use a very simple trader interface, similar to a web-trading interface of a typical online broker (see Figure \ref{fig:interface}).  Traders are allowed to only place market orders, and in order to elicit information, only the spot price is displayed. A trader can then offer a trade (buy or sell) and a desired quantity, upon which the trader is quoted a (volume weighted average) price. The trader may either accept or cancel the trade. 

\subsection{Experimental Design}

\begin{figure}[!h]
\begin{center}
\resizebox{0.35\textwidth}{!}{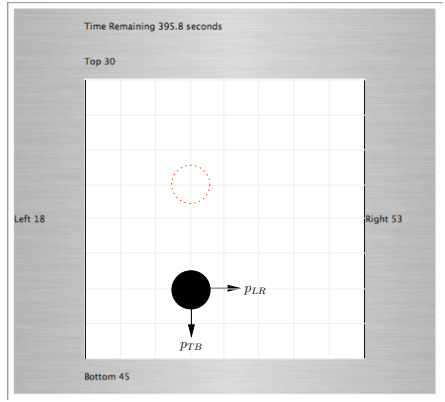}
\caption{2-dim random walk}
\label{fig:ExpDesign}
\end{center}
\end{figure}

There are two markets, \text{\sc lr} and {\sc tb}, which are based on the 
2 dimensional
random walk illustrated in Figure~\ref{fig:ExpDesign}.
The 2 dimensional random walk is two independent 1 dimensional random walks: horizontal (\text{\sc lr})
and vertical ({\sc tb}). Each random walk is a classic Gambler's Ruin problem \cite{feller1958introduction}. The starting position (indicated by the dotted  red circle)
 is 
\math{(x_0,y_0)}, and there are two probabilities, \math{p_{\text{\sc lr}}}, the probability of moving right in the horizontal dimension, and 
\math{p_{\text{\sc tb}}}, the probability of moving down in the vertical dimension. The random walk \math{(x,y)} is bounded in the grid
\math{[-S,S]^2}. So if \math{|x|\ge S}, the x-coordinate of the random
walk is restarted at \math{x_0} (the y-coordinate is left unchanged)
and similarly if  \math{|y|\ge S}, the y-coordinate of the random
walk is restarted at \math{y_0} (the x-coordinate is left unchanged).

The values of the markets \text{\sc lr} and {\sc tb} are defined before any particular experiment, based on how often the ball hits the right edge before the left edge, or the bottom edge before the top edge. The probability that the ball hits the right (resp. bottom)
edge before the left (resp. top) edge can be computed analytically \cite{feller1958introduction}.
In terms of \math{p_{\text{\sc lr}},\ p_{\text{\sc tb}},\ x_0,\ y_0,\ S} 
these values are (For \math{p\not=\frac12})\footnote{%
For \math{p=\frac12}
one has to take a limit (eg. \math{V_{\text{\sc lr}}=\frac{S+x_0}{2S}})}.
\mand{
V_{\text{\sc lr}}=
\frac
{\lambda_{\text{\sc lr}}^{S-x_0}-\lambda_{\text{\sc lr}}^{2S}}
{1-\lambda_{\text{\sc lr}}^{2S}},
\qquad
V_{\text{\sc tb}}=
\frac
{\lambda_{\text{\sc tb}}^{S-y_0}-\lambda_{\text{\sc tb}}^{2S}}
{1-\lambda_{\text{\sc tb}}^{2S}},
}
where \math{\lambda=p/(1-p)}.
Traders are allowed to simultaneously trade in both markets
{\sc lr} and {\sc tb}. For the experiments, we set 
\math{p_{\text{\sc lr}}=p_{\text{\sc tb}}=p} and
\math{x_0=y_0=z}. Thus, other than one market being visually represented
vertically and one horizontally, the two markets are completely symmetric.

\paragraph{Trader Signals}
The traders see a realization of the random walk unfolding over time.
As shown in Figure~\ref{fig:ExpDesign}, the number of times the walk has
 hit the left, right, top and bottom edges is shown, together with how much
time is left. 
A trader can estimate \math{V_{\text{\sc lr}}} and
\math{V_{\text{\sc tb}}} from these numbers; for example,
from the figure, we can make out from this partial realization of the 
random walk that 
\math{V_{\text{\sc lr}}\approx \frac{53}{71}=0.75} and
\math{V_{\text{\sc tb}}\approx \frac{45}{75}=0.60}. Although these are 
realizations of the same random process, we immediately see that the
trader is getting a noisy signal of the variable on the basis of which the market pays off
(as \math{t\rightarrow\infty}, traders would have perfect information that determines the payoff). 
This signal improves with time as more information is revealed; in particular,
in our example, the error in the traders signal decreases in proportion to 
\math{1/\sqrt{t}}. This process of gradual information revelation is similar to what
goes on in real prediction markets, with traders getting better information over time.

\paragraph{Market Shocks}
In a normal equilibrium setting the parameters \math{p,S,z} are fixed. We
can institute a market shock during the random walk by 
changing one or more of these parameters. Changing these parameters can reflect different types of market shocks in the real world -- for example, if $p$ changes, there is no visible cue, and traders have to infer a change in the underlying dynamics from observables.

\subsection{Description of Experiments}
Our data were collected in three distinct trading sessions; we used results from the first two sessions in order to improve the design of the second session. All the traders in our experiments were relatively sophisticated; they all had prior experience with the trading interface and knowledge of prediction markets. In each case they sat in the same room and traded using the web-based interface on their personal laptops.

The first two sessions were individual experiments in which students from a graduate-level Computational Finance class were recruited to participate. 11 and 9 students respectively chose to participate in the two experiments. Participants were incentivized with gift certificates: the trader with largest return received \$15; the second best received \$10 and the three next best traders \$5 each.

The third session was an educational deployment of the market as part of a graduate / advanced undergraduate class on E-Commerce. Students were studying prediction markets and participated in four trading games during the class. They were incentivized with the opportunity to earn extra credit in the class. 17 students participated. 10 points of total extra credit were allocated for the experiment, with the 10 points divided proportionally among all traders who overall made profit in the experiments. 

In each case, the LMSR based market maker was configured with the loss parameter $b$ set to 125. The information-based market maker was configured to begin with belief $\mu_t = 50$, $\sigma_t = 12$, and estimation of the trader noise given by $\sigma_\epsilon = 5$. The window of trade history for the adaptive mechanism was set to 5 for the first two experiments and to 10 for the last four. In the first two experiments traders started with $100,000$ units of currency and 0 shares, and were allowed to take both long and short positions in each market. In the four subsequent experiments, traders began with an initial endowment of $10,000$ units of currency, and $200$ shares, but were not permitted to take short positions. We flipped the market makers being used in the TB and LR markets between experiments.

We now provide details of the experiments. For experiments 1 and 2, traders were told that there may or may not be a change in the underlying parameters governing the random walk. The conditions for the remaining experiments are described below. In all cases except for Experiment 2, final payoffs were based on the analytically computed probabilities described above. A summary of the parameters used in each experiment are shown in Table \ref{tab:RWparams} below.

\textbf{Experiment 1: Equilibrium} Each trader viewed an independent realization of the random walk for 10 minutes, and so the traders had their own personal information set based on the random walk they were seeing, as well as the price dynamics which carried information regarding the realizations that other traders were seeing. 11 traders participated.

\textbf{Experiment 2: Common Information Shock} In this experiment, all traders viewed the same random walk realization, projected on a screen. They were told that the payoff of the markets would be the actual realized ratios of the two random walks, rather than the analytically computed probabilities. The parameters governing the random walk were ``shocked'' at the 5 minute mark. In this case, the traders' information gradually becomes completely correct, and the market maker is eventually trading against perfectly informed traders.

\textbf{Experiment 3: Limited Information Equilibrium} This experiment was similar to Experiment 1, except for the fact that traders only saw their personal realization of the random walk for 2 minutes. They were allowed to trade for 10 minutes. 17 students participated.

\textbf{Experiments 4, 5, and 6: Equilibrium With Probabilistic Shocks} Before these experiments, students were told that they would be participating in 3 consecutive games. In each of these games, the random walk would start off with some combination of parameters. With a 50\% chance, these parameters would change between minutes 3 and 7 of the random walk. Traders were not told whether or not there would be a jump in a particular experiment. A coin was flipped for each of the three experiments. There was no shock (change in parameters) in Experiments 4 and 5, while there was a shock in Experiment 6 (therefore we call it IndivInfoShock below). Trading went on for 10 minutes. 17 students participated.

\begin{table}
\begin{center}
\begin{tabular}{l|cccc}
&p&S&z&V\\ \hline
Equilibrium&0.600&4&-1&0.7322\\ \hline
CommonInfoShock&0.600&2&-1&0.5846\\
\hspace{.5in} jump to&0.600&2&+1&0.1231\\ \hline
LimitedInformation&0.764&4&-3&0.6912\\ \hline
Equilibrium(4)&0.533&4&-3&0.1897\\ \hline
Equilibrium(5)&0.866&4&-3&0.8453\\ \hline
IndivInfoShock&0.826&4&-3&0.7890\\
\hspace{.5in} jump to&0.516&4&-2&0.2999\\
\end{tabular}
\caption{Random walk parameters for each of the 6 human subject experiments.
\label{tab:RWparams}
}
\end{center}
\end{table}

\subsection{Results}

\begin{figure}
\begin{minipage}[c]{0.48\textwidth}
\begin{center}
\includegraphics*[width=2.3in]{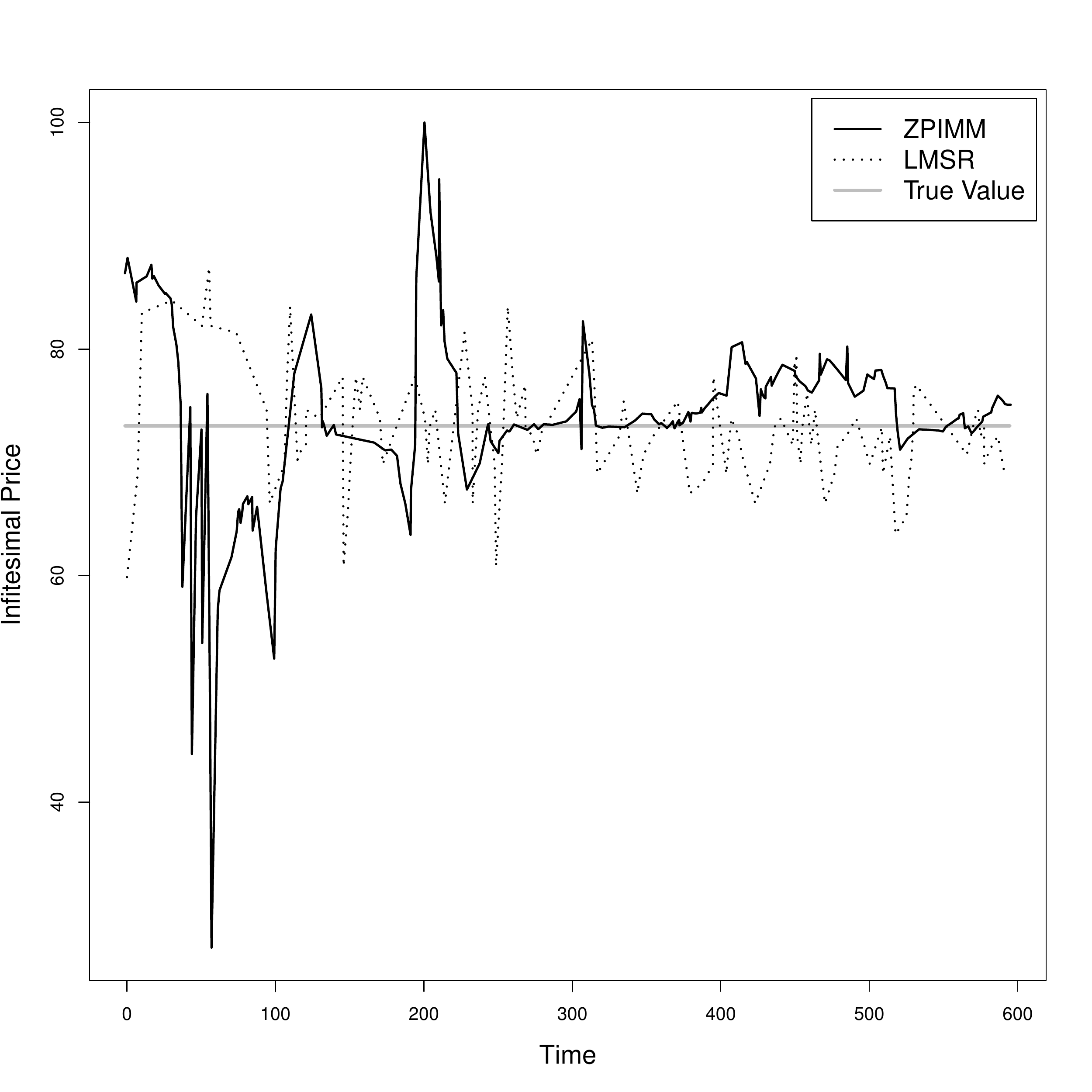}\\
(a) Equilibrium(1)\\
\includegraphics*[width=2.3in]{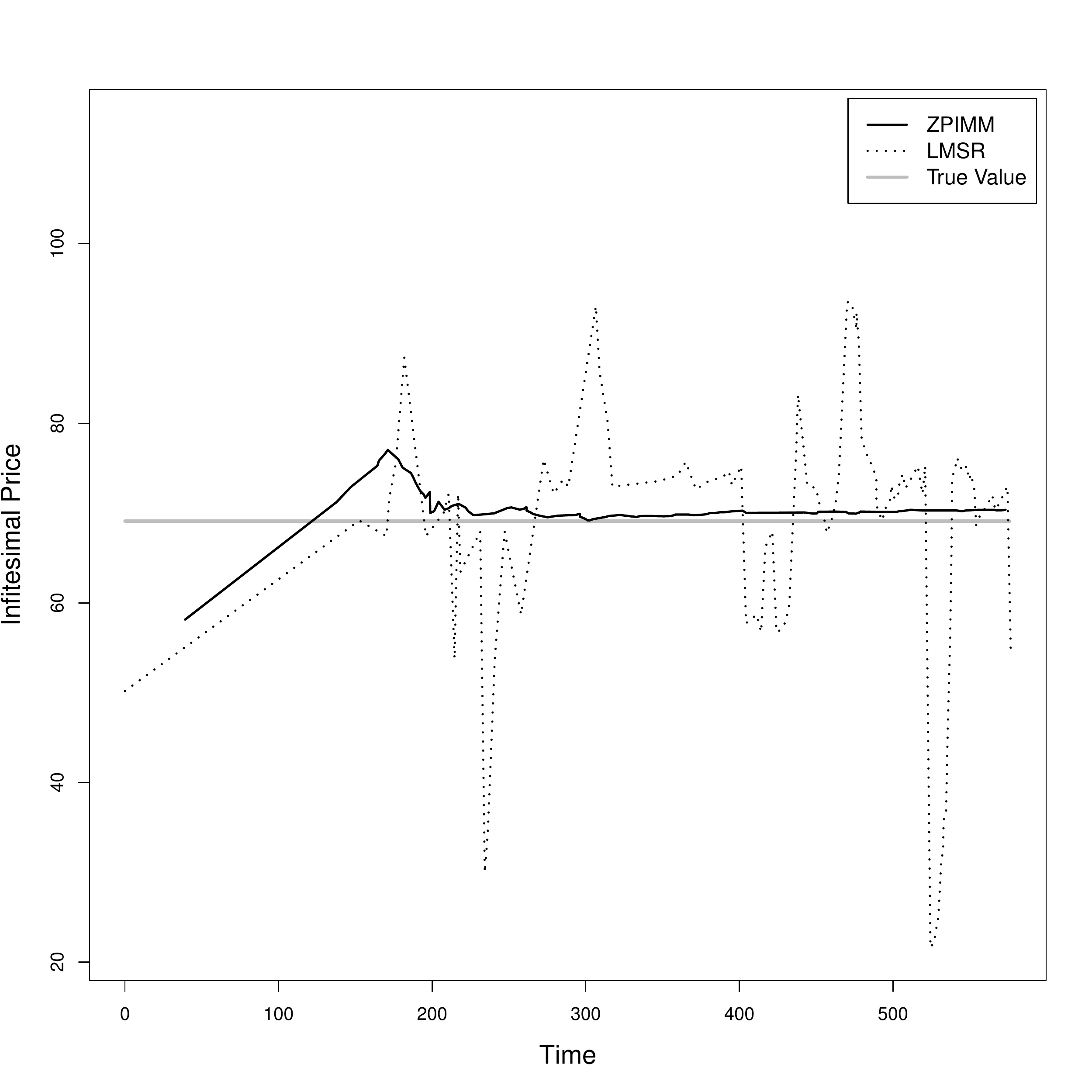}\\
(c) LimitedInformation\\
\includegraphics*[width=2.3in]{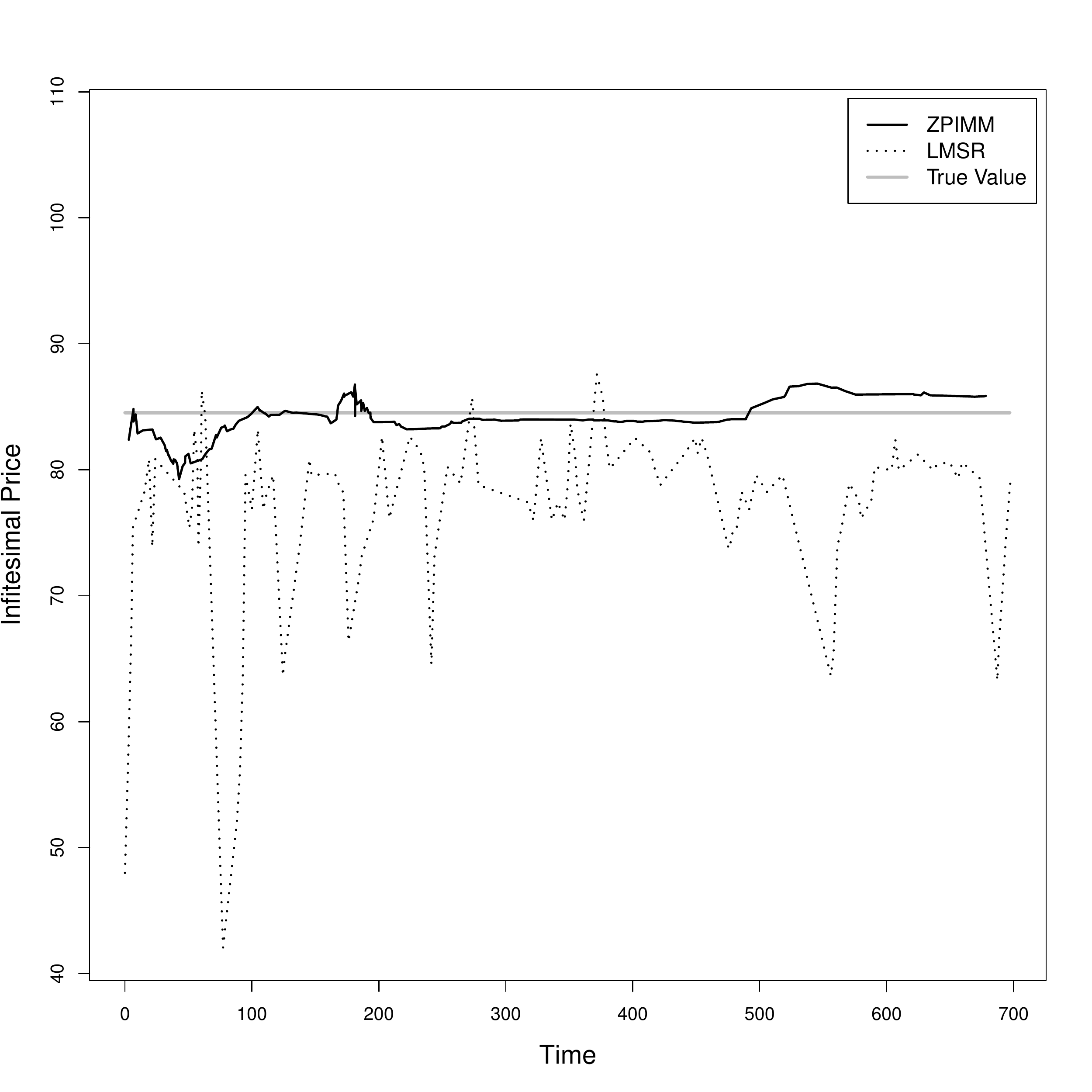}\\
(e) Equilibrium(5)
\end{center}
\end{minipage}
\begin{minipage}[c]{0.48\textwidth}
\begin{center}
\includegraphics*[width=2.3in]{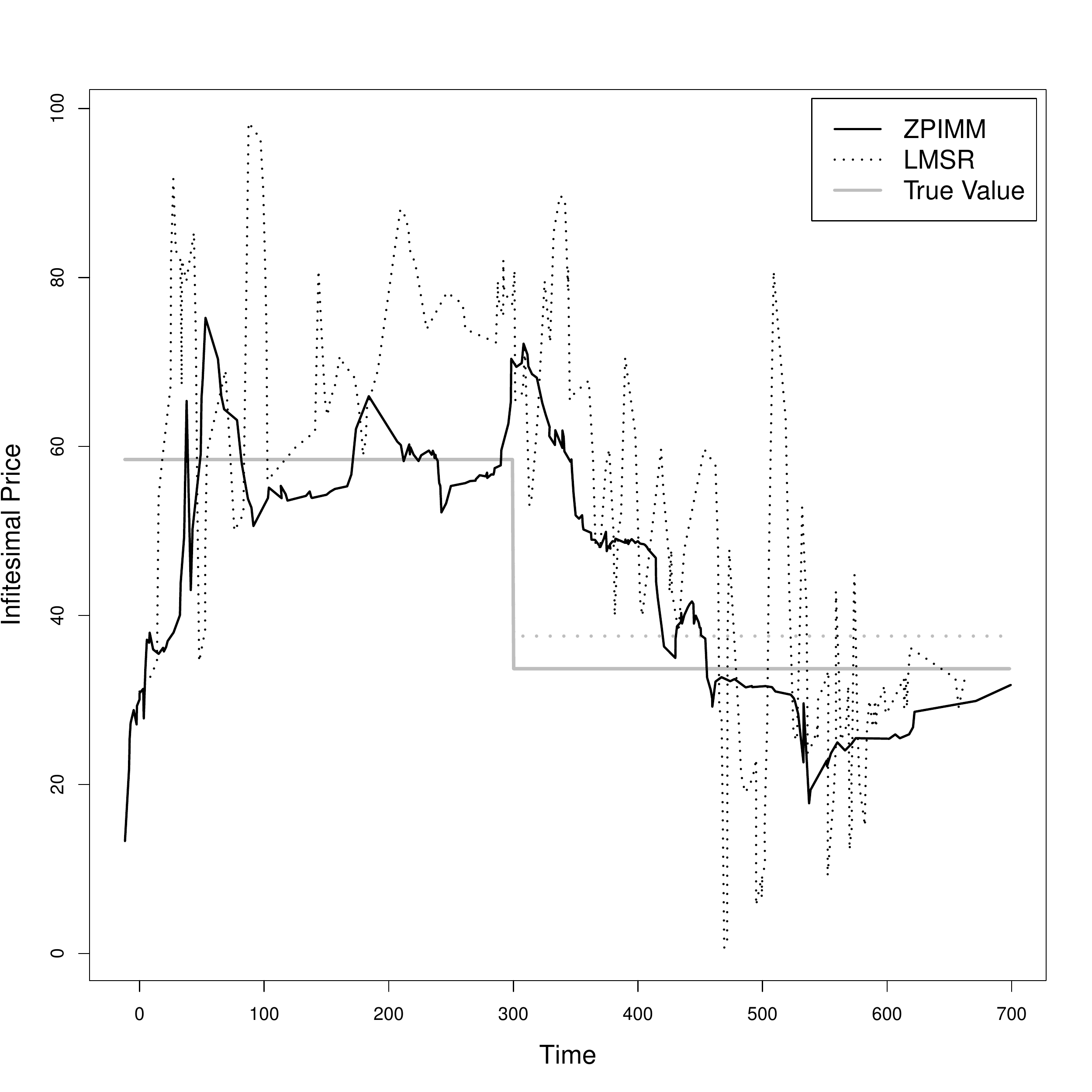}\\
(b) CommonInfoShock\\
\includegraphics*[width=2.3in]{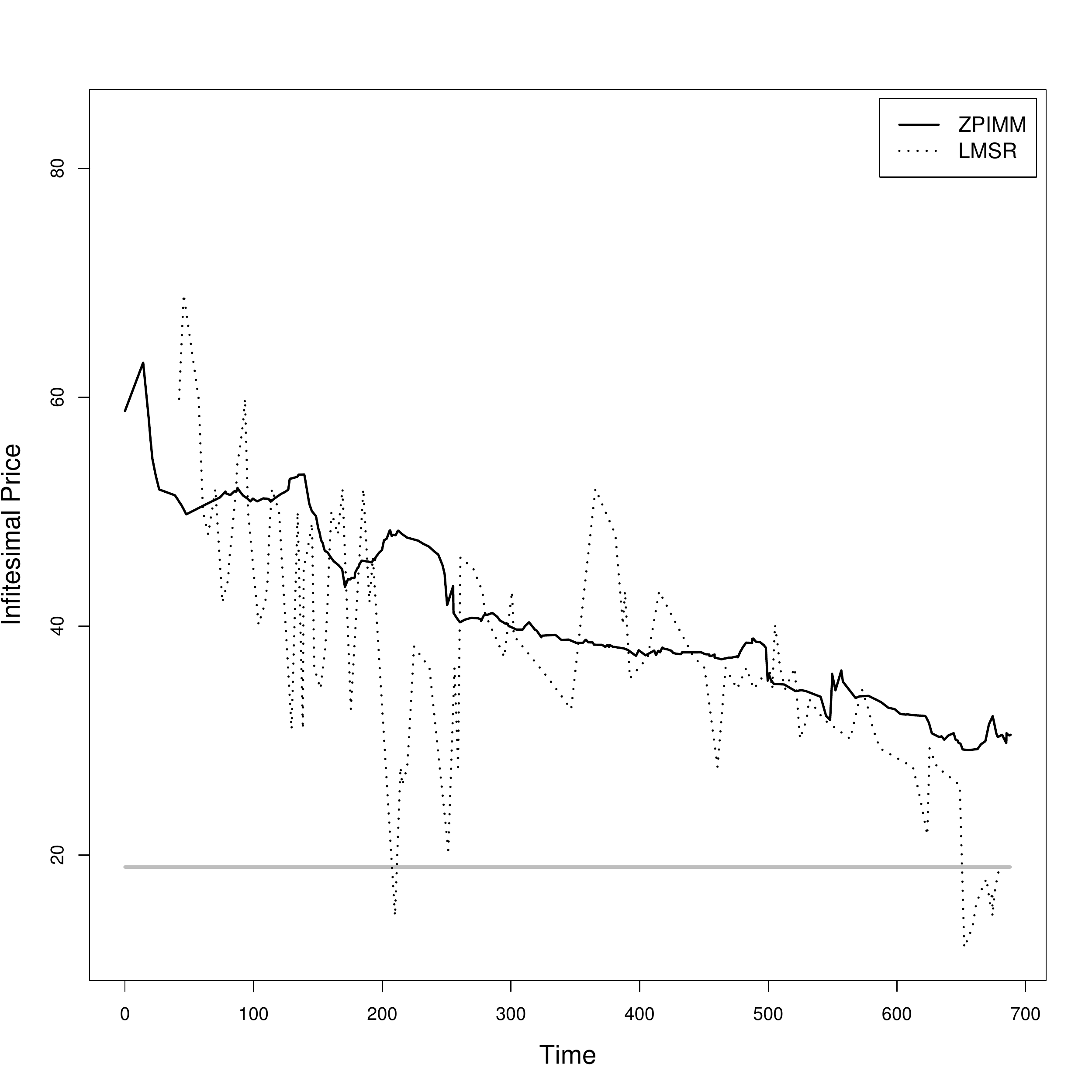}\\
(d) Equilibrium(4)\\
\includegraphics*[width=2.3in]{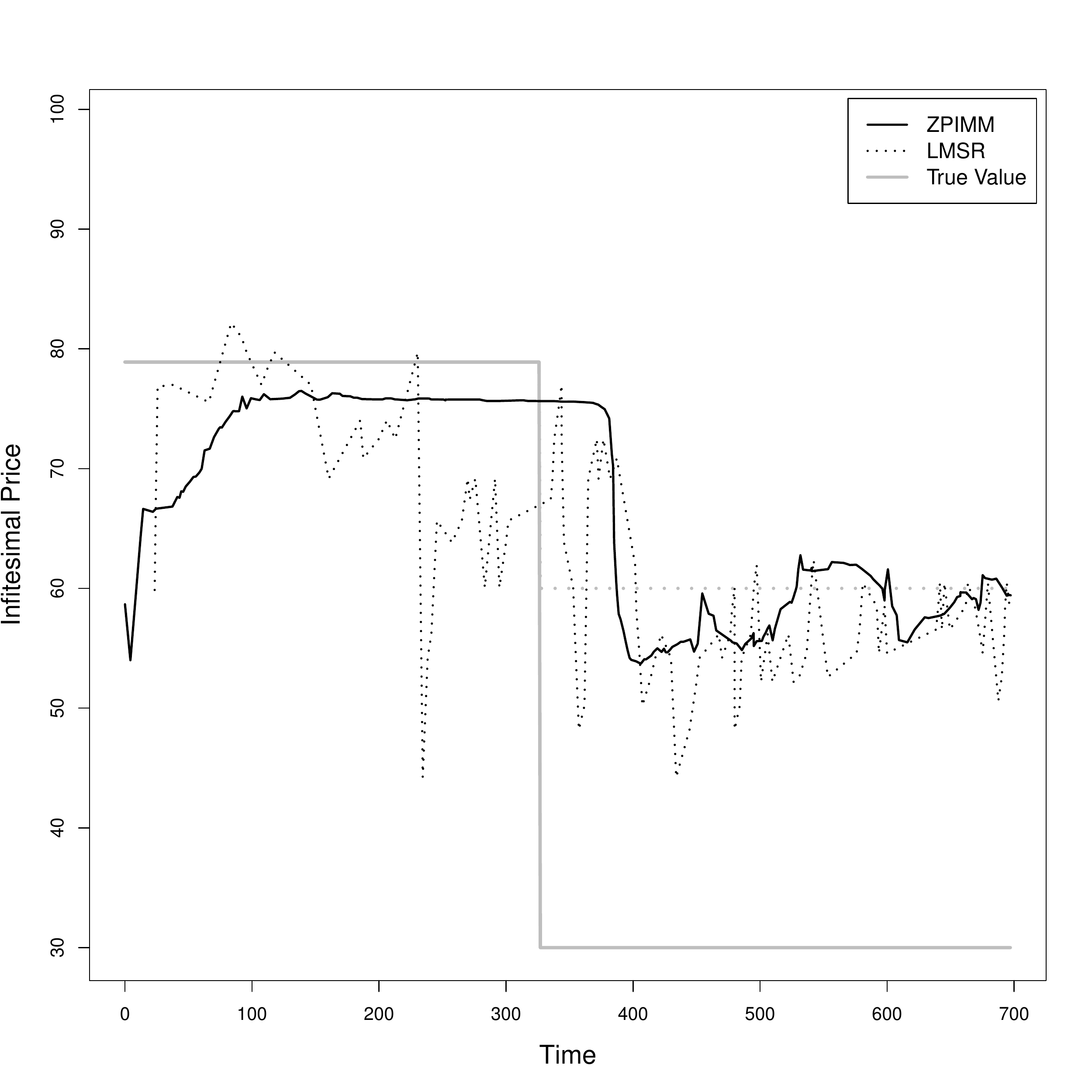}\\
(f) IndivInfoShock
\end{center}
\end{minipage}
\caption{Summary behavior of LMSR and BMM in live trading experiments. Experiments (a) and (b) are from one deployment in which BMM used a window size of 5 and (c) through (f) from a later one in which BMM used a window size of 10. In general, BMM exhibits more stable behavior than LMSR. Additionally, In the ``shock'' experiments (b) and (f), BMM is now able to adapt as well as, if not better than, LMSR. 
\label{fig:live}}
\end{figure}

Figure~\ref{fig:live} shows the main results of the experiments (the
analogs of the simulation data in Figure~\ref{fig:motiv}), and
Table \ref{tab:liveResults} shows some statistics on the price
processes. There are various interesting phenomena in the individual
experiments, discussed below, but the big picture is relatively
clear. BMM dominates LMSR in terms of profit made in five of the six
live experiments, while at the same time producing a more stable price
process, with better price discovery, as measured by distance from the
``true'' value (RMSD values in Table~\ref{tab:liveResults}. The
behavior of BMM is improved in experiments 3 through 6, which were run
with a longer adaptive window of 10, leading to more stability (and
potentially slower adaptivity). While higher values of the $b$
parameter for LMSR would lead to improved stability (and slower
adaptivity), this would come at the cost of making even greater
losses.

\begin{table}
\begin{center}
\begin{tabular}{l|ll|ll|ll|ll|}
&\multicolumn{2}{c|}{Profit} & \multicolumn{2}{c|}{Spread} & \multicolumn{2}{c|}{RMSD} & \multicolumn{2}{c|}{RMSDeq} \\
& LMSR & BMM & LMSR & BMM & LMSR & BMM & LMSR & BMM \\ \hline
Equilibrium(1)&-1350.12&47231.77&3.12&4.04&4.92&7.66&3.73&3.01\\
CommonInfoShock&-1510.89&8972.50&3.06&3.21&20.67&15.98&16.51&6.76\\
LimitedInformation&-1602.14&4083.95&1.61&0.49&14.43&2.15&14.56&0.93\\
Equilibrium(4)&-2619.07&-10588.86&1.81&0.95 &21.67&23.13&14.82&17.05\\
Equilibrium(5)&-3168.55&9134.58&1.42&0.51 &11.18&1.5&8.15&1\\
IndivInfoShock&-92.29&20226.44&1.89&0.89&8.87&6.47&6.88&6.6
\end{tabular}
\end{center}
\caption{Summary statistics of MM performance in the live trading experiments. 
Spreads are the offered spread for 40 shares (the average trade size
observed) at each point in time when a trade executed. RMSD is the root mean square deviation of the MM's belief from the true value. $RMSD_{eq}$ is the same metric evaluated after ``convergence,'' defined for convenience as the time period halfway between the last change to the true value and the end of the trading period.
\label{tab:liveResults}
}
\end{table}

\paragraph{Experiments 1 and 2, and lessons learned}

In the Equilibrium(1) experiment, there are some severe fluctuations
at about the 75 sec and 200 sec marks. The fluctuations around the 75
sec mark are probably due to individuals who had outlier realizations
early on. The fluctuations around the 200 sec mark are due to a single
irrational ``rogue'' trader who was willing to buy at a price of
100. Unfortunately, since there is no penalty for random wild trading
(unlike in real financial markets), such behavior is bound to arise
with human experiments. Discounting these anomalous trades, BMM
converges quite nicely to equilibrium, as does LMSR (except for its
characteristic oscillations).  Further, in the MarketShock experiment,
BMM now adapts as fast if not faster than Hanson. 

The BMM profit in the Equilibrium(1) experiment is a little misleading
because about 30,000 of it was due to the rogue trader; BMM does what
it is supposed to do though, by adapting and making profit based on
its Bayesian learned valuation. This wild trader also accounts for the
increased RMSD of BMM in this experiment. After the market
equilibrates and finds the true value, the RMSD of BMM dominates LMSR
(the \math{\text{RMSD}_{\text{eq}}} row in
Table~\ref{tab:liveResults}). Similarly, when the market is close to
equilibrium in the MarketShock experiment (in this case, after seven
and a half minutes of trading time in total; since the jump occurs
after five minutes, we give the market half of the remaining time to
equilibrate) the RMSD of BMM dominates LMSR by a significant margin.

These experiments reveal a couple of interesting facts. First, the
behavior of some rogue traders can seriously impact outcomes. In this
case, it seems that, when given large initial endowments and the
ability to sell short, some traders use their market power to full
effect without worrying about profit. So we decided to give people
more ``reasonable'' endowments in the future, including an endowment
of stock to start with, and prohibit short-selling. This likely leads to a
psychologically more understandable scenario for participants, and
less possibility for arbitrary manipulation by traders who are
psychologically uninvested in the outcome. 

\begin{figure}
\begin{minipage}[c]{0.48\textwidth}
\begin{center}
\includegraphics*[width=2.3in]{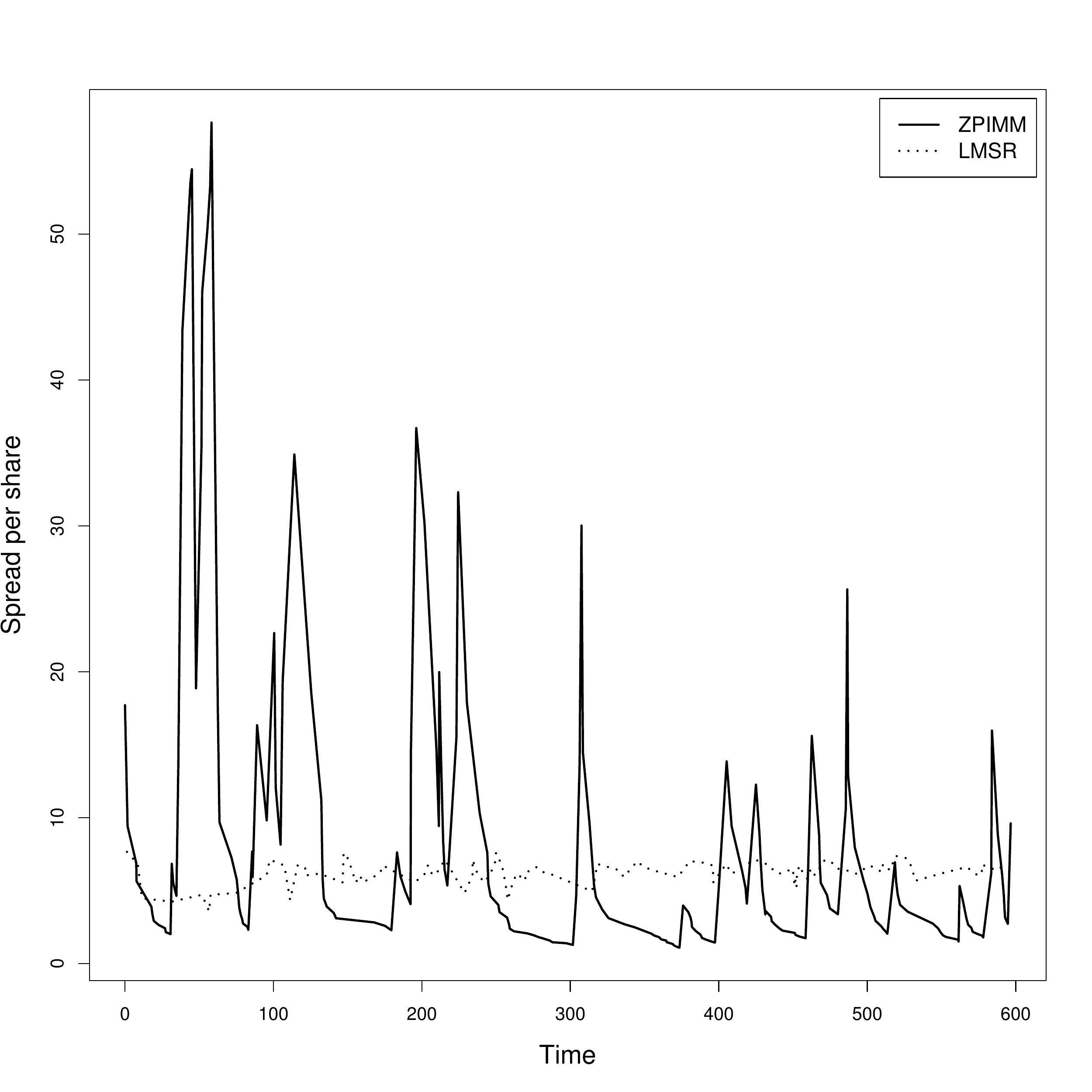}\\
(a) Equilibrium(1)
\end{center}
\end{minipage}
\begin{minipage}[c]{0.48\textwidth}
\begin{center}
\includegraphics*[width=2.3in]{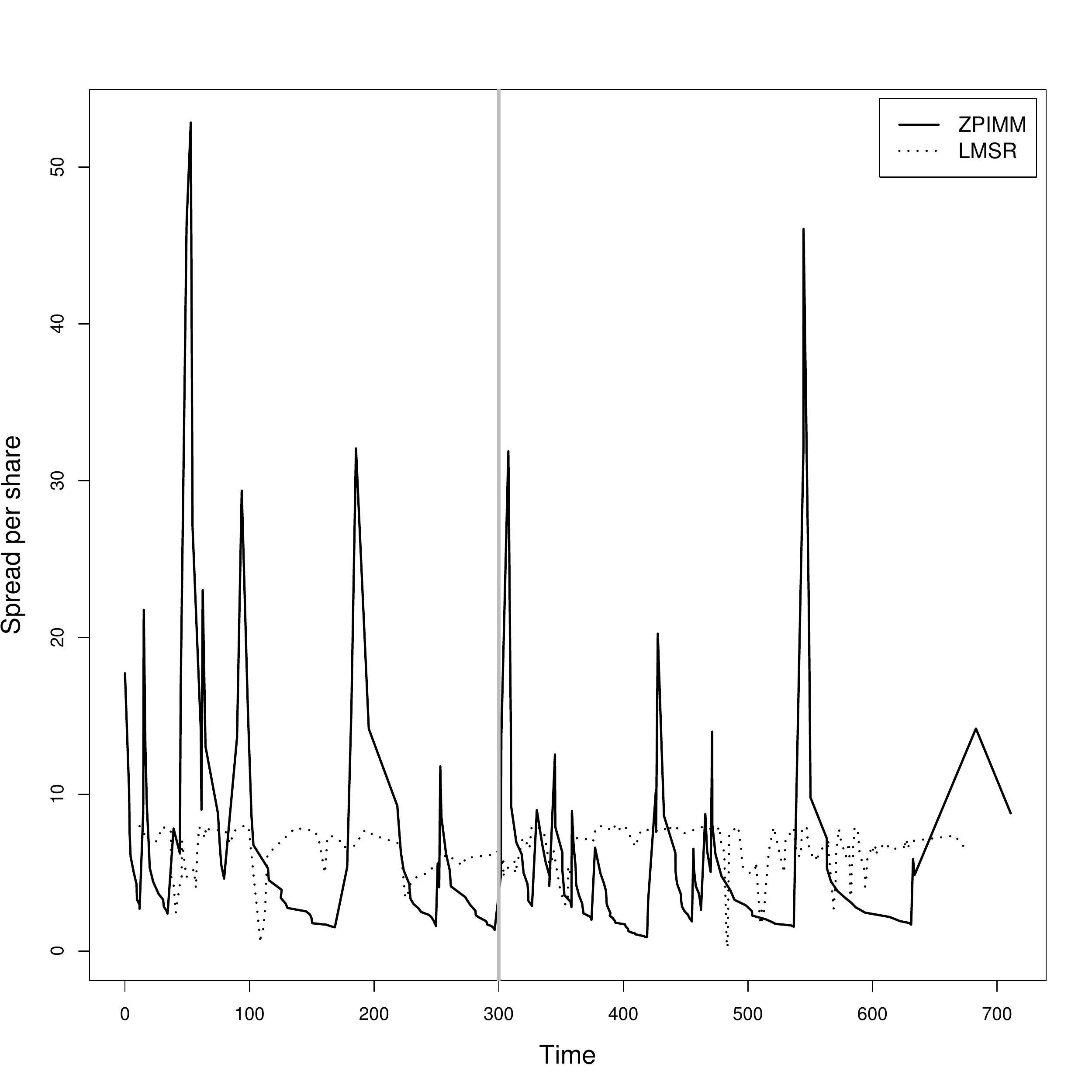}\\
(b) CommonInfoShock
\end{center}
\end{minipage}
\caption{Dynamic behavior of LMSR and BMM spread in Experiments 1 and
  2. BMM often increases its spread in response to unlikely trade
  sequences, making it highly adaptive, while sacrificing some stability.
\label{fig:spreadsPt1}}
\end{figure}

Second, the spreads and behavior of BMM were somewhat less stable than
we had expected based on simulation. Figure \ref{fig:spreadsPt1} shows
that BMM often increases the spread in response to market conditions,
even though there are relatively few shocks in the system. While this
still yields good behavior, we hypothesized that tweaking the window
parameter would lead to more stable behavior without sacrificing
adaptivity too much. Therefore, we changed the window size to 10 for
the next set of experiments.

\paragraph{Experiments 3 through 6}

Experiments 3 through 6 demonstrate the typical behaviors of BMM and
LMSR clearly. There are a couple of interesting details that emerge
from the experiments. First, in Experiment 4 (Equilibrium(4)),
convergence to the true value is very slow for both LMSR and
BMM. While LMSR comes close to the true value in the last few seconds
before the end of trading, BMM fails to do so. We hypothesize that
this is because this market was the only one in which the true value
was below the starting value of 50, and thus necessitated people
selling their initial endowment to get to the true value. In this
case, BMM also takes a fairly substantial loss, because it was misled
by the trading behavior.  

Second, in the IndivInfoShock experiment (number 6), while traders
were told that the true value would only be the true value \emph{after
the shock}, later interviews with participants revealed that they
thought the true value would be the average of the two true
values. Therefore, the stock ended up trading at around 60, instead of
the final true value of 30. 

\begin{figure}
\begin{minipage}[c]{0.48\textwidth}
\begin{center}
\includegraphics*[width=2.3in]{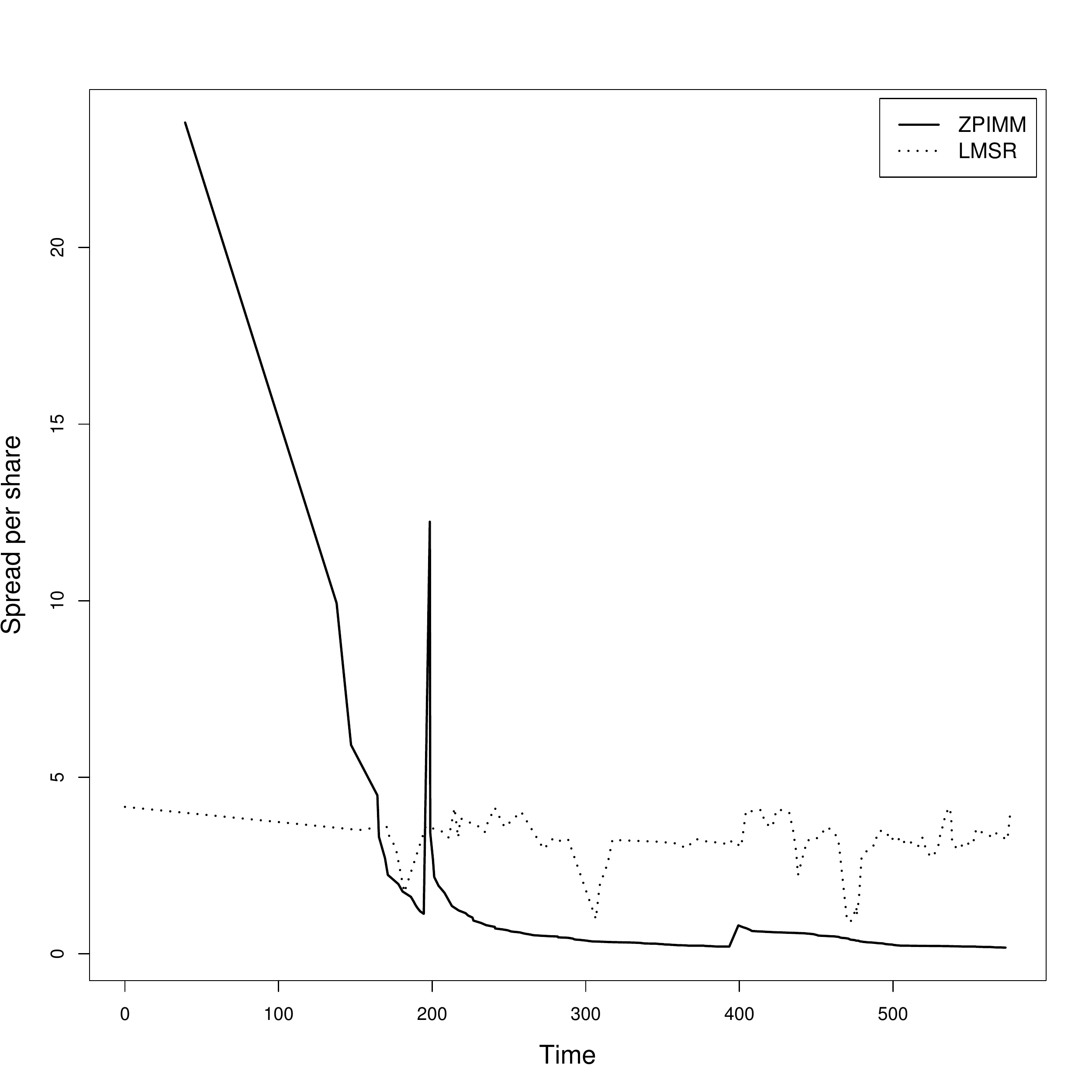}\\
(a) LimitedInformation\\
\includegraphics*[width=2.3in]{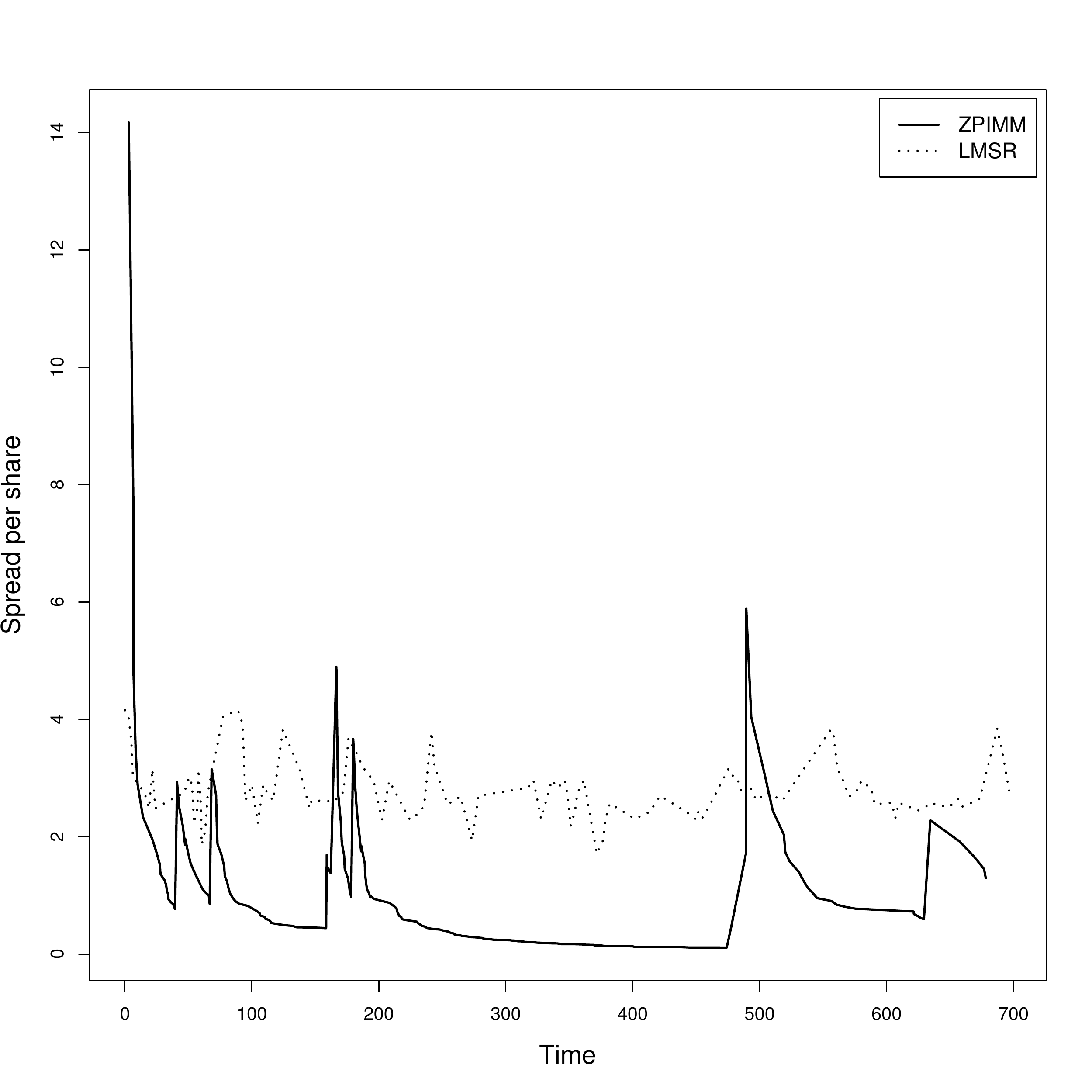}\\
(c) Equilibrium(5)
\end{center}
\end{minipage}
\begin{minipage}[c]{0.48\textwidth}
\begin{center}
\includegraphics*[width=2.3in]{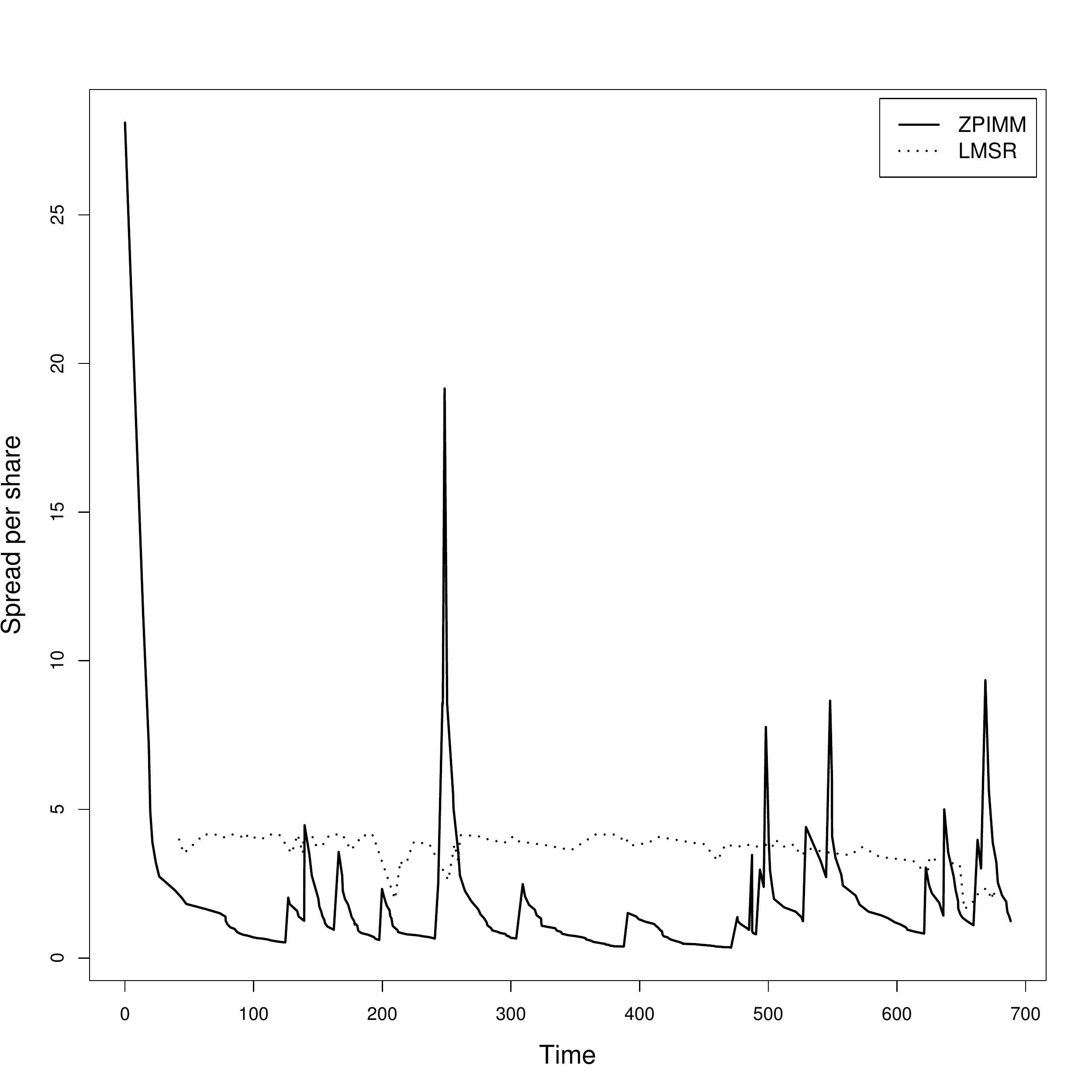}\\
(b) Equilibrium(4)\\
\includegraphics*[width=2.3in]{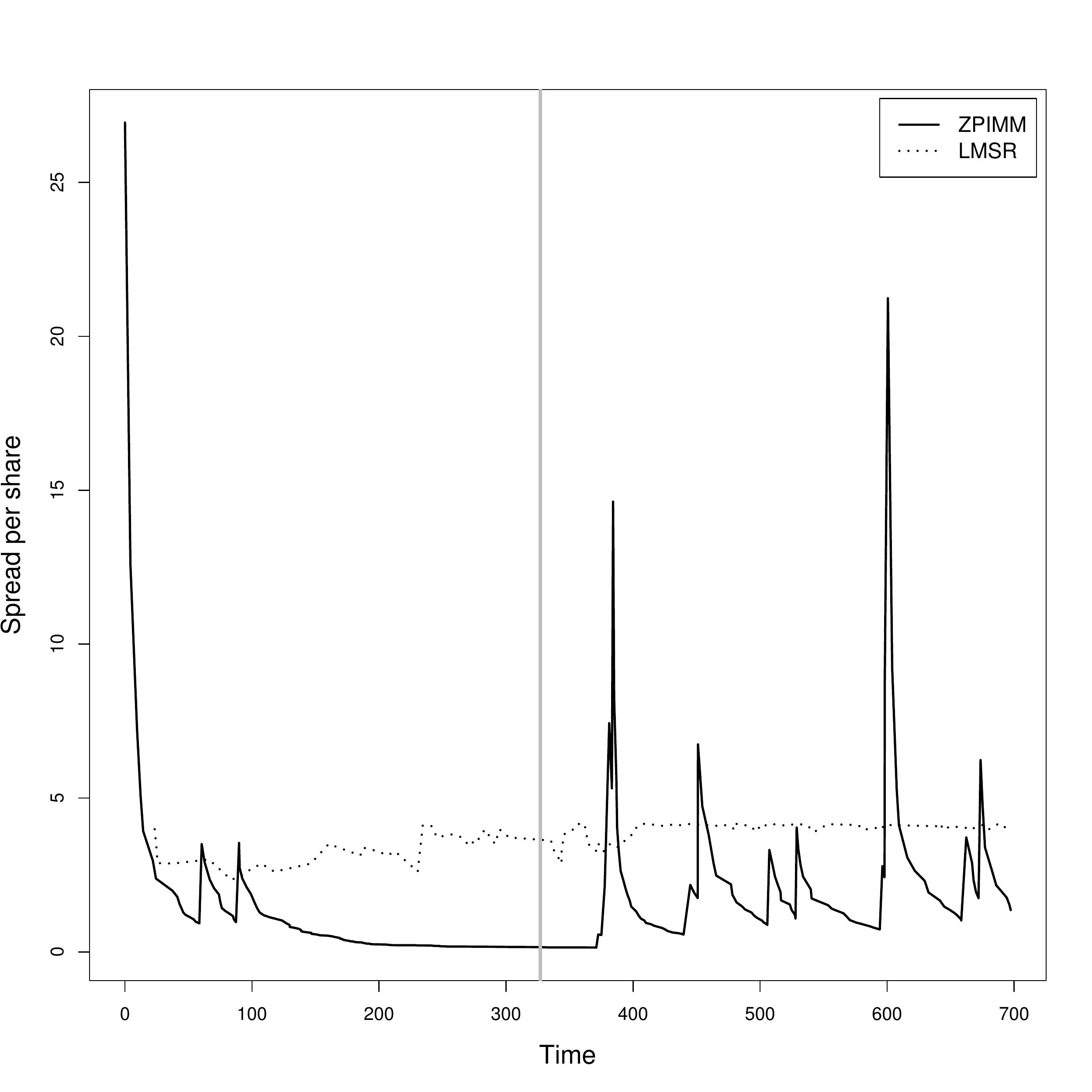}\\
(d) IndivInfoShock
\end{center}
\end{minipage}
\caption{Dynamic behavior of LMSR and BMM spread in Experiments 3
  through 6. BMM is now significantly more stable, but (d) shows it is
  still quick to adapt.
\label{fig:spreadsPt2}}
\end{figure}

In both these cases, it is nice that the symmetry of the experimental
design enables fair comparison between LMSR and ZPIMM: trader behavior
leads to anomalies for \emph{both} market makers. Figure
\ref{fig:spreadsPt2} shows the behavior of the spreads for the two
market makers, demonstrating that BMM is significantly more stable in
these experiments, while still reasonably adaptive, as evidenced by
Experiment 6 (IndivInfoShock).

\paragraph{Number of trades}
The table below provides a summary of the comparison of the LMSR and
BMM market makers in terms of the number of \emph{confirmed} buy or
sell trades executed in each experiment. The numbers are roughly
comparable, although LMSR usually comes in a little bit higher,
presumably because it provides profit opportunities due to price
fluctuations when BMM has stabilized.

\begin{center}
\begin{tabular}{l|cc|cc|}
&\multicolumn{2}{c|}{\# Buy/Sell}&\# Traders\\
&LMSR&BMM\\ \hline
Equilibrium&119&146&11\\
CommonInfoShock&213&131&9\\
LimitedInformation&116&63&17\\
Equilibrium(2)&109&113&17\\
Equilibrium(3)&118&85&17\\
IndivInfoShock&113&121&17
\end{tabular}
\end{center}

\paragraph{Summary}

Our live trading experiments demonstrate several key facts. First,
while LMSR has nice theoretical properties that suggest it will
converge to rational expectations equilibria, in practice this is
asking an awful lot of the participants in the market. As long as
traders' posterior beliefs do not converge to a single point, there
will remain trading incentives, and this is in evidence in all our
experiments. LMSR suffers from characteristic fluctuations in the spot
price even after it should have attained equilibrium. BMM, on the
other hand, provides a tighter belief once it has converged, and has
attractive potential to make markets without losing money, or even at
a profit. It manages this while providing superior price discovery and
spread properties in our live trading experiments.

Experiment 4 (Equilibrium(4)) provides evidence that BMM may sometimes
suffer high losses, especially when the market behaves
strangely. While occassional such instances are not a huge problem, it
will be important to monitor and understand the circumstances that can
lead to high losses so that we can ensure that they cannot be
reproduced by manipulators intentionally deceiving the market maker.




\section{Conclusion}
\label{section:concl}
We have presented an adaptive, information based Bayesian
market maker BMM. In simulation as well as in live trading, when controlling for liquidity (as measured by spread), BMM demonstrates significantly better convergent behavior at equilibrium than Hanson's LMSR market maker, while being equally adaptive to changes in the market's valuation of the security. BMM also provides a meaningful quantitative 
posterior probability distribution for the value of the security being traded. Further, it does not lose money (in fact it typically makes money) in both the idealized simulated markets and live trading, implying that it could provide substantially better liquidity at lower cost than Hanson's LMSR. The caveat is that, unlike Hanson's LMSR market maker, BMM is not loss 
bounded. BMM thus provides a real alternative to
Hanson's LMSR for market making in real information markets, with many potential
benefits. 

Our second goal was to present a symmetric, fair experimental design
for comparing two market makers in a live trading setting. We have only
begun to explore the possibilities of this paradigm, which offers
a realistic trading environment in which traders gradually get information 
as they trade against the dealer. The design allows one to 
study market shocks with and without visible cues, to study convergence and
adaptation of market makers, and to study real trader behavior -- 
how do traders really trade given their valuation and the market price?

There has been recent recognition in the literature of some of the
problems with LMSR; in particular, Othman \emph{et al} have noted its
liquidity-insensitivity and the difficulty of setting the $b$
parameter appropriately \cite{DBLP:conf/sigecom/OthmanSPR10}. They
have proposed an alternative approach which varies the $b$ parameter
-- this is another potential alternative to LMSR that is important to
explore and characterize, and our experimental platform provides a
good method for testing this market maker as well.

It is worth noting that all the experiments described in this paper
are information \emph{aggregation} experiments. Many prediction
markets are important because of their information
\emph{dissemination} role (see for example, Othman and Sandholm's
recent work on a market to predict the opening of the new Computer
Science building at Carnegie-Mellon University
\cite{DBLP:conf/sigecom/OthmanS10}). In these cases, only one or a few
insiders have knowledge of the true value, and the market's goal is to
incentivize these insiders to reveal their information. Testing market
makers in such settings is important.

Future work on developing BMM further should focus on characterizing
situations where it can potentially make significant losses, and
attempt to mitigate losses in these situations. In particular, BMM's
susceptibility to manipulation from a trader who understands and
attempts to beat the algorithm is not yet well understood. Further
experimental evaluation is also necessary. Another interesting open
question is whether we can formulate precisely the tradeoff between
convergence, loss and adaptability for market makers.

\bibliographystyle{plain}
\bibliography{predmarkets}  

\end{document}